\documentclass[12pt,onecolumn,journal]{IEEEtran}

\usepackage[export]{adjustbox}
\usepackage{setspace}
\usepackage{bbm}
\usepackage{amsmath,graphics,amssymb,epsfig,subfigure,color,amsthm,cite}
\usepackage{array}
\usepackage{multirow}
\usepackage{enumerate}
\usepackage[ruled]{algorithm2e}
\SetKwInOut{Input}{Input}
\SetKwInOut{Output}{Output}
\DontPrintSemicolon

\usepackage{bm}
\usepackage[cmintegrals]{newtxmath}
\usepackage{tablefootnote}

\theoremstyle{definition}
\newtheorem{theorem}{Theorem}

\newcommand{\rom}[1]{\uppercase\expandafter{\romannumeral #1 \relax}}
\newtheorem{prop}{Proposition}

\begin{document}
 \title{Sparse Joint Transmission for Cloud Radio Access Networks with Limited Fronthaul Capacity} 

\author{\IEEEauthorblockN{Deokhwan Han,~\IEEEmembership{Student Member,~IEEE}, \IEEEauthorblockN{Jeonghun Park},~\IEEEmembership{Member,~IEEE},\\ Seok-Hwan Park,~\IEEEmembership{Senior Member,~IEEE}, and \IEEEauthorblockN{Namyoon~Lee},~\IEEEmembership{Senior Member,~IEEE}}\\
 \thanks{D. Han and N. Lee are with the Department of Electrical Engineering, POSTECH, Pohang, Gyeongbuk 37673, South Korea  (e-mail: \{dhhan, nylee\}@postech.ac.kr).}
\thanks{J. Park is with the School of Electronics Engineering, College of IT Engineering, Kyungpook National University, Daegu, 41566, South Korea (e-mail: jeonghun.park@knu.ac.kr).}
\thanks{S.-H. Park is with the Division of Electronics Engineering, College of Engineering, Jeonbuk National University, Jeonju, 54896, South Korea (e-mail: seokhwan@jbnu.ac.kr).}
\thanks{This work was partly supported by Institute of Information \& communications Technology Planning \& Evaluation(IITP) (No.2021-0-00161, Post MIMO system research for massive connectivity and new wireless spectrum) and by the National Research Foundation of Korea (NRF) grant funded by the Korea government (MSIT) (No. 2020R1C1C1013381).}}

 \maketitle
\begin{abstract}


A cloud radio access network (C-RAN) is a promising cellular network, wherein densely deployed multi-antenna remote-radio-heads (RRHs) jointly serve many users using the same time-frequency resource. By extremely high signaling overheads for both channel state information (CSI) acquisition and data sharing at a baseband unit (BBU), finding a joint transmission strategy with a significantly reduced signaling overhead is indispensable to achieve the cooperation gain in practical C-RANs. In this paper, we present a novel sparse joint transmission (sparse-JT) method for C-RANs, where the number of transmit antennas per unit area is much larger than the active downlink user density. Considering the effects of noisy-and-incomplete CSI and the quantization errors in data sharing by a finite-rate fronthaul capacity, the key innovation of sparse-JT is to find a joint solution for cooperative RRH clusters, beamforming vectors, and power allocation to maximize a lower bound of the sum-spectral efficiency under the sparsity constraint of active RRHs. To find such a solution, we present a computationally efficient algorithm that guarantees to find a local-optimal solution for a relaxed sum-spectral efficiency maximization problem. By system-level simulations, we exhibit that sparse-JT provides significant gains in ergodic spectral efficiencies compared to existing joint transmissions.
  
\end{abstract}

\IEEEpeerreviewmaketitle
\section{Introduction}
\subsection{Motivation}

Next-generation cellular networks, including 6G, require to support demands on high speed and uniform data services \cite{boccardi2014five}. The ever-growing demands for higher bit rates and more uniform data services necessitate novel cellular network architectures that can yield high network capacity within a limited spectrum. The new cellular architectures providing an increased network capacity are expected to have two key ingredients: 1) densely deployed base stations (BSs) topologies that aggressively reuse spectrum \cite{ge20165g} and 2) the coordination among the BSs to eliminate both inter-user and inter-cell interference \cite{gesbert2010multi,lozano2013fundamental,lee2014spectral}.

 A cloud radio access network (C-RAN) \cite{chen2011c,boccardi2014five,gesbert2010multi,lozano2013fundamental,lee2014spectral} is a promising  cellular architecture to achieve high energy and spectral efficiencies by both network densification and BS cooperation gains. A cloud-RAN consists of distributed antennas, called remote radio heads (RRHs), connected to a centralized baseband unit (BBU) pool via high-speed fronthaul links. This network virtually forms a large-scale distributed and cooperative MIMO system. The centralized BBU pool can jointly perform user selection, beamforming, and power allocation for both downlink and uplink communications to eliminate interference between scheduled users. As the network density increases, this joint processing allows to achieve a high cell-splitting gain by reducing the communication distance between the network and the mobile users; thereby, it can significantly dwindle the transmission power.  
 
Unfortunately, in practice, the promising gain by the joint transmission comes at the cost of prohibitively high signaling overhead. Specifically, for downlink communications, BBU needs to acquire global channel state information (CSI) and to share the precoded data with RRHs.  As the network becomes denser, the amount of signaling overheads for CSI acquisition and data sharing increases tremendously.  Moreover,  acquiring global CSI perfectly and sharing the precoded data without any error is impossible due to a finite-rate fronthaul capacity. For instance, in C-RAN operating with time-division-duplexing (TDD) mode, each RRH estimates users' channels via uplink pilots and sends them to BBU via a finite-rate fronthaul link. Therefore, the accuracy of CSI at BBU is fundamentally limited by both channel estimation errors and the fronthaul capacity. Furthermore, the precoded data symbols at BBU are shared with RRHs through finite-rate fronthaul links for the downlink transmission. A low-rate fronthaul link introduces a high quantization error on the downlink data; this leads to the degradation of the downlink performance. Considering the signaling overheads and limited fronthaul capacity constraints, the effective gain of the joint processing offered by C-RANs can be very marginal. 

To enhance the effective gain in practice, the joint transmission exploiting a few dynamically selected RRHs is a promising solution because it can considerably reduce the signaling overheads associated with the joint processing. For example, from the users' viewpoint, it is better to receive the downlink signals from all RRHs to increase data rates. Whereas, from the network perspective, the use of all RRHs increases the associated signaling overheads for joint transmission. In particular, when the active user density is much smaller than the total number of antennas per unit area in the network, the use of sparsely chosen RRHs would be sufficient for joint transmission, while it considerably reduces the overheads.  In this sense, it is essential to use a sparse RRH cooperation method to form a large-scale C-RAN. Unfortunately, finding the jointly optimal solution for the sparsely chosen cooperative RRH sets per user, precoding vectors, and transmit power, which maximizes the downlink sum-spectral efficiency, is a well-known NP-hard problem \cite{luo2008dynamic, yu2013multicell, hong2013joint}, even under assumptions of the perfect and global CSI and the infinite-rate fronthaul capacity. Considering the practical constraints of a finite-rate capacity of fronthaul links and noisy and partial CSI, finding a local-optimal solution for the sum-spectral efficiency maximization problem is highly non-trivial.  To tackle this problem, this paper introduces a novel sparse joint downlink transmission technique that maximizes a lower bound of the sum-spectral efficiency under practical constraints.

\subsection{Related Works}

%
 
The joint transmission by a sparsely chosen set of RRHs is proposed as an energy-efficient solution for downlink transmissions of C-RANs \cite{shi2014group,dai2016energy,shi2018enhanced,huang2015joint}. The common approach is to design the network-wide sparse precoding vector to minimize a total number of active RRHs (equivalently network-wide power consumption) under a set of user rate constraints \cite{shi2014group,dai2016energy,shi2018enhanced,huang2015joint}. Specifically, a novel group-sparsity beamforming framework is presented in \cite{shi2014group}, in which the weighted $\ell_1$ and $\ell_2$-norm minimization techniques are taken to promote the group sparsity using a successive convex approximation technique. In \cite{dai2016energy}, an efficient group-sparsity beamforming algorithm is introduced by using the reweighted $\ell_1$ minimization \cite{candes2008enhancing}. In \cite{huang2015joint}, a two-stage algorithm is presented, in which the set of active RRHs is initially identified in a user-centric manner, and BBU designs joint precoding vectors for the chosen RRH set to mitigate the inter-user interference. However, these prior studies focused on the precoding design to minimize the total transmission power rather than sum-spectral efficiency maximization. Therefore, it is unclear how the sum-spectral efficiency behaves as the number of active RRHs becomes sparse in the network.

The sparse-beamforming algorithm is also proposed to maximize the sum-spectral efficiency under limited fronthaul capacity \cite{dai2014sparse}. This algorithm uses both the generalized weighted minimum mean squares error (WMMSE) technique in \cite{christensen2008weighted}, and the reweighted $\ell_1$ minimization method \cite{candes2008enhancing} to find the beamforming solution under a finite-rate fronthaul link constraint. These studies, however, assume perfect and global CSI at BBU, thereby it cannot reflect the effects of channel estimation and fronthaul quantization errors in practical systems. In addition, by the nature of the WMMSE optimization framework, the computational complexity to implement the sparse-beamforming algorithm in \cite{dai2014sparse} is the order of $\mathcal{O}\left((KLN)^{3.5}\right)$ per iteration, where $K$, $L$, and $N$ are the number of users, RRHs, and the number of antennas per RRH, respectively. The high computational complexity hinders to use the WMMSE method for large-scale C-RAN systems.  

Another popular approach to reducing the signaling overheads for the joint transmission is to exploit edge-computing capabilities with local caches \cite{tao2016content,peng2017layered,peng2014joint}. For instance, content-centric sparse multicast beamforming is proposed in \cite{tao2016content}, where users who request the same content are clustered and apply the sparse multicast precoding using local caches at each RRHs. In addition, a three-stage layered group sparse beamforming (LGSBF) algorithm \cite{peng2017layered} is introduced to obtain a joint solution of adaptive RRH selection, backhaul content assignment, and multicast beamforming.  Although these studies show the benefits of the content-based clustering and transmission in reducing the signaling overheads associated with the joint transmission,  they require additional resources such as local caches at RRHs, which is a different assumption from our work.

The most relevant prior work from the viewpoint of the optimization framework is \cite{choi2019joint}. In contrast to \cite{choi2019joint}, in which CSI sharing is only assumed for coordinated beamforming, in this paper, we consider both data and CSI sharing for joint transmission by incorporating the quantization error effects by limited fronthaul capacity. In addition, we also consider sparse joint transmission unlike \cite{choi2019joint}. The block sparsity constraint imposed by the sparse joint transmission yields a unique challenge in the design of the precoding algorithm compared to the algorithm in \cite{choi2019joint}. The first-order optimality condition differs from that in \cite{choi2019joint}; thereby, our algorithm finding the stationary point is distinct from the algorithm introduced in \cite{choi2019joint},  albeit they share a generalized power iteration principle.

\subsection{Contributions}
This paper considers a joint RRH clustering, beamforming, and power optimization problem for downlink  C-RAN. The main contributions of this paper are summarized as follows:

\begin{itemize}

 \item We derive a lower bound expression of a downlink sum-spectral efficiency for C-RAN considering effects of noisy CSI and quantization error in data sharing by finite-rate fronthaul links. In particular, using the notion of generalized mutual information \cite{yoo2006capacity,medard2000effect,lapidoth2002fading,ding2010maximum}, we establish a lower bound expression as a function of relevant system parameters, including channel estimation error and a finite-rate fronthaul capacity. 
    
   \item We propose a unified optimization framework that finds the network-wide sparse precoding vector to maximize the lower bound of sum-spectral efficiency. Unlike the WMMSE optimization framework, \cite{dai2014sparse}, the key innovation is to convert the sum-spectral efficiency maximization problem under the sparsely cooperative RRHs constraint into a tractable non-convex optimization by mapping all optimization variables into a high dimensional space using the recently developed large-scale optimization techniques in \cite{choi2018joint,choi2019joint,han2020distributed}. The tractable non-convex optimization is the form of maximizing the product of Rayleigh quotients under the sparse RRH activation constraint. This formulation can be regarded as a generalized sparse principal component analysis (sparse-PCA) problem.  By relaxing the sparse active RRH constraint into a non-convex function, we formulate a unified non-convex optimization problem that finds the network-wide sparse precoding vector while reducing the quantization errors to maximize the spectral efficiency. 
 

 \item  We derive the local optimality conditions for the reformulated non-convex optimization problem. To accomplish this, we characterize the first- and the second-order necessary conditions for the local optimality. In particular, we derive a condition in a closed-form to verify that a saddle point can be a local optimum. 

  \item Using the derived optimality conditions, we present a sparse joint transmission algorithm that jointly identifies a set of active RRHs, the precoding vectors (for beamforming and compression), and the power allocation for RRHs.  The sparse joint transmission (sparse-JT) algorithm guarantees to find a local-optimal solution for the reformulated non-convex optimization problem. Besides, the computational complexity of the proposed algorithm increases linearly with the number of downlink users, quadratically with both the number of RRHs $L$, and the antennas per RRH $N$.  This complexity implies that the proposed algorithm is scalable to use  C-RANs.

\item We show numerically that the proposed sparse joint transmission algorithm considerably outperforms the existing user-centric RRH clustering with WMMSE and zero-forcing (ZF) precoding methods in different CSI and fronthaul link capacity conditions.  This confirms that sparse-JT can achieve a higher synergetic gain of clustering and precoding than the existing methods in C-RANs.
\end{itemize}

%
%

\section{System Model}

We consider a  C-RAN network where $L$ RRHs, each equipped with $N$ antennas, jointly send downlink signals to $K$ single-antenna users.  We assume that the $\ell$th RRH is connected to a BBU via fronthaul links with a finite-rate $C_{\ell}$ bits per second. Each RRH has a transmit power budget $P$.



\subsection{Noisy-and-Incomplete Downlink CSIT Acquisition}
\begin{figure}[t]
	\centering
    \includegraphics[width=10cm]{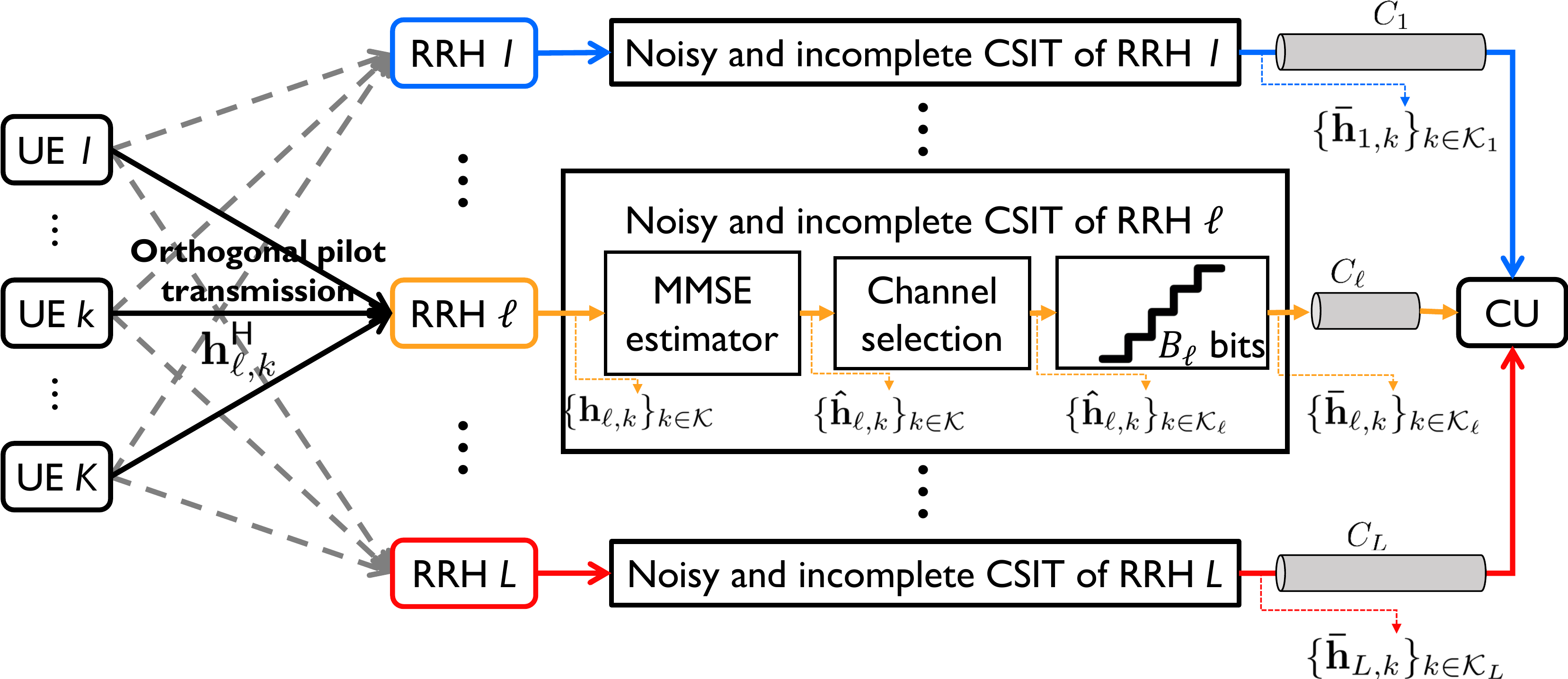}
  \caption{An illustration of noisy-and-incomplete CSI acquisition model.  In our model, each RRH estimates the noisy channels, selects a few strongest channels, and quantizes the selected channels to meet the finite-rate fronthaul capacity. } \label{Fig_UL}
\end{figure}

We present a noisy downlink CSIT acquisition model as shown in Fig. \ref{Fig_UL}.  Let ${\bf h}_{\ell,k}=\left[h_{\ell,k}^1,\ldots, h_{\ell,k}^N\right]^{\sf T}$ be the downlink channel vector from the $\ell$th RRH to the $k$th user. This channel vector is modeled as \begin{align}
  {\bf h}_{\ell,k} = \beta_{\ell,k}^{1/2}{{\bf g}}_{\ell,k} \in \mathbb{C}^{N\times 1},   
\end{align}
where $\beta_{\ell,k}\in\mathbb{R}$ and ${\bf g}_{\ell,k}\in\mathbb{C}^{N\times 1}$ are a large-scale fading coefficient and a small-scale fading vector, respectively. The distribution of ${\bf g}_{\ell,k}$ is assumed to be the complex Gaussian, i.e., ${\bf g}_{\ell,k}\sim \mathcal{CN}\left({\bf 0},{\bf R}_{\ell,k}\right)$, where ${\bf R}_{\ell,k}=\mathbb{E}\left[{\bf g}_{\ell,k}{\bf g}_{\ell,k}^{\sf H}\right]\in\mathbb{C}^{N\times N}$ is the spatial covariance matrix of the channel. 

{\bf MMSE channel estimation per RRH:} Thanks to channel reciprocity in TDD mode, the $\ell$th  RRH estimates downlink channel ${\bf h}_{\ell,k}$ by estimating the uplink channel vector ${\bf h}_{\ell,k}^{\sf H}$. Under the premise that each user sends orthogonal pilot sequences with length $\tau \ge K$, the minimum mean square error (MMSE) estimation of ${\bf h}_{\ell,k}$, i.e., ${\bf\hat h}_{\ell,k}=\left[{\hat h}_{\ell,k}^1,\ldots, {\hat h}_{\ell,k}^N\right]^{\sf T}$, is given by
\begin{align}
    {\bf\hat h}_{\ell,k} = {\bf h}_{\ell,k} + {\bf e}_{\ell,k},
    \label{eq:Imperfect_CSI}
\end{align}
where ${\bf e}_{\ell,k}=\left[e_{\ell,k}^1,\ldots, e_{\ell,k}^N\right]^{\sf T}$ is the estimation error vector. Assuming the Gaussian noise in the channel estimation, ${\bf e}_{\ell,k}$ is distributed by zero-mean Gaussian with covariance matrix ${\bf \Phi}_{\ell,k} = \mathbb{E}\left[{\bf e}_{\ell,k}{\bf e}_{\ell,k}^{\sf H}\right]\in\mathbb{C}^{N\times N}$, and it is statistically independent of $ {\bf\hat h}_{\ell,k}$. 
Assuming that $p^{{\sf ul}}$ is the uplink pilot transmission power, the channel estimation error covariance matrix is given as a function of spatial covariance matrix ${\bf R}_{\ell,k}$, large-scale fading coefficient $\beta_{\ell,k}$, pilot length $\tau$, and pilot transmission power $p^{{\sf ul}}$ \cite{hoydis2011massive,yin2013coordinated}:
\begin{align}
 {\bf \Phi}_{\ell,k}    &= {\beta}_{\ell,k}{\bf R}_{\ell,k}-{\beta}_{\ell,k}^2{\bf R}_{\ell,k}\left({\beta}_{\ell,k}{\bf R}_{\ell,k}+\frac{\sigma^2}{\tau p^{\sf ul}}{\bf I}_{N}\right)^{-1}{\bf R}_{{\ell},k}.
\end{align}

{\bf Channel selection:} We present two channel selection methods using 1) instantaneous CSI and 2) average received signal power at the RRHs. First, using the MMSE channel estimator, RRH $\ell\in \mathcal{L}$ has knowledge of noisy versions of channel vectors, i.e., $\{{\bf \hat h}_{\ell,1},\hdots,{\bf \hat h}_{\ell,K}\}$. Sending all estimated channel vectors perfectly from the RRH to BBU is infeasible under a finite-rate fronthaul constraint. To compress CSI information, we consider a simple channel selection method. The key idea is to choose the best $U_{\ell}(\leq K)$ channel vectors in the order of the channel gains.  Let ${\bf \hat h}_{\ell,\pi_{\ell}(k)}$ be the estimated channel vector of the $\ell$th RRH with the $k$th largest channel gain, where $\pi_{\ell}(k)\in \mathcal{K}=\{1,\hdots,K\}$ be the index function such that $\|{\bf \hat h}_{\ell,\pi_{\ell}(1)}\|_2^2\geq \|{\bf \hat h}_{\ell,\pi_{\ell}(2)}\|_2^2\geq \cdots \geq \|{\bf \hat h}_{\ell,\pi_{\ell}(K)}\|_2^2 $. Then, each RRH sends the top-$U_{\ell}$ channel vectors, i.e., $\left\{{\bf \hat h}_{\ell,k} \right\}$ for $k\in \mathcal{K}_{\ell}=\{\pi_{\ell}(1),\ldots, \pi_{\ell}(U_{\ell})\}$ to BBU, where $U_{\ell}$ is chosen as a function of the fronthaul link capacity $C_{\ell}$. For instance, the fronthaul capacity is extremely limited, RRH $\ell$ can select $U_{\ell}=1$, implying that the best user channel only is sent to BBU.  For ease of explanation, we define a subset RRHs that has knowledge of the channel vector for user $k$ by $\mathcal{L}_k=\{\ell~|~k\in\mathcal{K}_{\ell},\forall\ell\}\subset\mathcal{L}$. This index set will be used in the sequel.

In addition, to reduce the CSI acquisition overhead, we propose a simple strategy that estimates the channels for a few strongest channel links in the received power at RRHs. Specifically, each RRH periodically measures the uplink received power of all users, and selects $U_{\ell}(\leq K)$ users in the order of received power at RRH $\ell$ for $\ell\in \mathcal{L}$. Then, RRHs perform the channel estimation to acquire CSI for the selected users, and send the limited CSI to the BBU to generate a precoding solution. To validate the effect of this limited CSI acquisition strategy, we compare the ergodic sum-spectral efficiency performance with the case of using full CSI acquisition at RRHs in Section \rom{6}.

{\bf Channel quantization:} The selected estimated channel, $\left\{{\bf\hat h}_{\ell,\pi_{\ell}(1)},\hdots,{\bf\hat h}_{\ell,\pi_{\ell}(U_{\ell})}\right\}$, is quantized by using a simple uniform scalar (element-wise) quantizer with $B_{\ell}$ bits resolution. Then, the quantized CSI is sent to BBU via a finite rate fronthaul link $C_{\ell}$ bits per channel use. We assume that the quantization is performed independently across different antennas per RRH, ${{\hat h}}_{\ell,k}^n$ and ${{\hat h}}_{\ell,k}^m$ for $n \neq m$. This element-wise uniform quantization method is not optimal because it ignores the statistical correlation effect among the channel coefficients across antennas and RRHs  \cite{park2013joint,park2014fronthaul}. Nevertheless, we ignore the spatial correlation effects in the quantization error because their impacts are negligible when using a few-bit quantizer, and we shall focus on this quantization technique because it is more practically relevant from an implementation perspective.

 Using standard rate-distortion theory \cite{mezghani2007modified,zhang2016mixed,gersho2012vector},  we model the quantization process for the estimated channel of the $n$th antenna at the $\ell$th RRH as
\begin{align}
    {\bar h}_{\ell,k}^n &= {\hat h}_{\ell,k}^n + q_{\ell,k}^n, \quad\forall k \in \mathcal{K}_{\ell},\label{eq:Quantization_Error}
\end{align}
where $q_{\ell,k}^n$ is the quantization noise of ${\hat h}_{\ell,k}^n$ which is assumed to be the complex Gaussian with zero-mean and variance $\mathbb{E}[|q_{\ell,k}^n|^2 ]={\sigma}^2_{q_{\ell,k}^n}$, i.e., $q_{\ell,k}^n\sim\mathcal{CN}\left({ 0},{\sigma}^2_{q_{\ell,k}^n}\right)$. When using the uniform scalar quantizer with $B_{\ell}\ge6$ bits, it has shown in \cite{mezghani2007modified,gersho2012vector} that the variance of quantization noise is tightly approximated as
\begin{align}
 {\sigma}^2_{q_{\ell,k}^n} &\simeq \frac{\pi\sqrt{3}}{2}2^{-2B_{\ell}} \mathbb{E}\left[|\hat{h}_{\ell,k}^n|^2\right]
	\nonumber \\
	&= \frac{\pi\sqrt{3}}{2}2^{-2B_{\ell}}
	\left[
	{\beta}_{\ell,k}^2{\bf R}_{\ell,k}\left({\beta}_{\ell,k}{\bf R}_{\ell,k}+\frac{\sigma^2}{\tau p^{\sf ul}}{\bf I}_{N}\right)^{-1}{\bf R}_{{\ell},k}
	\right]_{n,n}. \label{eq:quantizaton_noise}
\end{align}
Therefore, the quantized signals $\{{\bar h}_{\ell,\pi_{\ell}(1)}^1,\hdots,{\bar h}_{\ell,\pi_{\ell}(1)}^N,\hdots,{\bar h}_{\ell,\pi_{\ell}(U_{\ell})}^N\}$, each with $2B_{\ell}$ bits, are reliably delivered from BBU to the $\ell$th RRH with the rate of 

\begin{align}
	\sum_{k\in\mathcal{K}_{\ell}}\sum_{n=1}^NI({\bar h}_{\ell,k}^n;{\hat h}_{\ell,k}^n)
	&=\sum_{k\in\mathcal{K}_{\ell}}\sum_{n=1}^N\log_2\left(1+\frac{\mathbb{E}\left[|\hat{h}_{\ell,k}^n|^2\right]}{{\sigma}^2_{q_{\ell,k}^n}}\right)\nonumber\\
	&\simeq U_{\ell}N\log_2\left(1+\frac{2}{\pi \sqrt{3}}2^{2B_{\ell}}\right), \label{eq:backhaul_capacity_constraint}
\end{align}
where the first equality follows from the differential entropy of complex Gaussian random variables ${\bar h}_{\ell,k}^n\sim \mathcal{CN}\left(0,\mathbb{E}\left[|\hat{h}_{\ell,k}^n|^2\right]+{\sigma}^2_{q_{\ell,k}^n}\right)$ and ${\hat h}_{\ell,k}^n\sim \mathcal{CN}\left(0,{\sigma}^2_{q_{\ell,k}^n}\right)$, and the second approximation holds from ${\sigma}^2_{q_{\ell,k}^n} \simeq \frac{\pi\sqrt{3}}{2}2^{-2B_{\ell}} \mathbb{E}\left[|\hat{h}_{\ell,k}^n|^2\right]$ in \eqref{eq:quantizaton_noise}.  When $B_{\ell}$ is sufficiently large, i.e., $B_{\ell}\ge3$, it boils down to
\begin{align}
	U_{\ell}N\log_2\left(1+\frac{2}{\pi \sqrt{3}}2^{2B_{\ell}}\right)&\simeq U_{\ell}N\log_2\left(\frac{2}{\pi \sqrt{3}}2^{2B_{\ell}}\right)\nonumber\\
	&=U_{\ell}N\left\{\log_2\left(2^{2B_{\ell}}\right)-\log_2\left(\frac{\pi \sqrt{3}}{2}\right)\right\}\nonumber\\
	&=U_{\ell}N\left(2B_{\ell}-1.444\right).\label{eq:HighApproximation}
\end{align}

Assuming the equal quantization bit allocation strategy per antenna, RRH requires to select the maximum number of quantization bits $B$ to minimize ${\sigma}^2_{{q}_{\ell,k}^{n}}$, while ensuring the fronthaul capacity constraint of $U_{\ell}N\log_2\left(1+\frac{2}{\pi \sqrt{3}}2^{2B_{\ell}}\right)\leq C_{\ell}$. This condition leads to the choice of the number of quantization bits per fronthaul link \begin{align}
	B_{\ell}^{\star}=\left\lfloor\frac{1}{2}\log_2\left(\frac{\pi\sqrt{3}}{2}\left(2^{\frac{C_{\ell}}{U_{\ell}N}}-1\right)\right)\right\rfloor. \label{eq:optimal_B}
\end{align}
When we denote ${\bf q}_{\ell,k}=\left[q_{\ell,k}^1,\ldots, q_{\ell,k}^N\right]^{\sf T}$, the covariance matrix becomes $\mathbb{E}\left[{\bf q}_{\ell,k}{\bf q}_{\ell,k}^{\sf H}\right]= {\bf Q}_{\ell,k}(B_{\ell})\simeq {\sf diag}\left( {\sigma}^2_{{q}_{\ell,k}^{1}} ,\ldots,  {\sigma}^2_{{q}_{\ell,k}^{N}} \right)$. By setting $B_{\ell}^{\star}$ as in \eqref{eq:optimal_B}, it is possible to meet the fronthaul capacity constraints for a given number of antennas and selected users $U_{\ell}$. If the quantization bit $B_{\ell}$ is fixed, to satisfy the constraint, one may alternatively choose $U_{\ell}^{\star}$ such that
\begin{align}
    U_{\ell}^{\star}=\left\lfloor\frac{C_{\ell}}{N\log_2\left(1+\frac{2}{\pi\sqrt{3}}2^{2B_{\ell}}\right)}\right\rfloor.
\end{align}
As a result, our CSI compression strategy, including the channel selection and quantization, can meet the fronthaul capacity constraint $C_{\ell}$ by flexibly choosing both the number of selected channels to share $U_{\ell}$ and the number of quantization bits to represent each selected channel values $B_{\ell}$.  The effect of the trade-off between $U_{\ell}$ and $B_{\ell}$ for given $C_{\ell}$ will be shown numerically in the simulation section.

\subsection{Downlink Transmission with Limited Fronthaul Capacity}

Using the proposed CSI estimation and compression strategy, BBU has noisy-and-incomplete CSIT $\{{\bf\bar h}_{\ell,k}\}_{\ell\in\mathcal{L},k\in\mathcal{K}_{\ell}}$. This subsection explains how BBU performs joint precoding to send downlink data symbols using this partial downlink channel knowledge.


{\bf Linear precoding:} Let $s_k[t]$ and ${\bf f}_{\ell,k}$ be a downlink transmit symbol to user $k$ in the $t$th time slot and the linear precoding vector being used at the $\ell$th RRH to deliver $s_k[t]$. When the coherence time interval is given by $\tau_c$, we assume that $s_k[t]$ is drawn from a complex Gaussian codebook with the average power $P=\mathbb{E}\left[|s_k[t]|^2\right]$ in the $t$th time slot where $t\in[\tau_c]$. Then, the precoded complex downlink signal of RRH $\ell$ is represented by a linear superposition of precoder ${{\bf f}_{\ell,k}}s_k[t]$ for $k\in\mathcal{K}$, i.e., 
\begin{align}
  {\bf x}_{\ell}[t] = \sum_{k\in\mathcal{K}} {\bf f}_{\ell,k} s_{k}[t],~~~~ \forall\ell\in \mathcal{L}.
\end{align}
\begin{figure}[t]
	\centering
    \includegraphics[width=10cm]{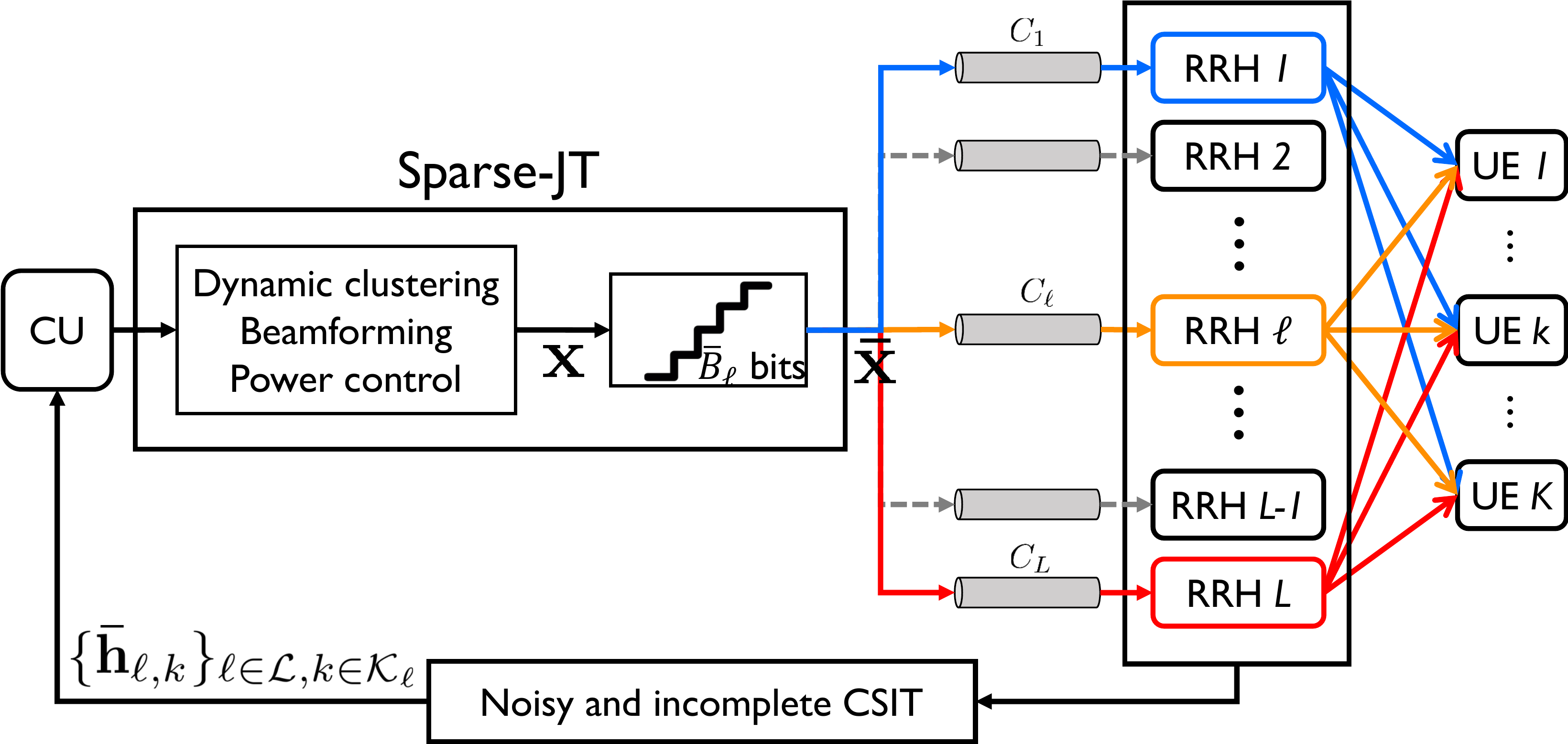}
  \caption{The proposed sparse-JT for C-RANs.} \label{Fig_DL}
\end{figure}

{\bf Precoded signal quantization:} In the similar manner of the channel quantization process, the precoded signal ${\bf x}_{\ell}[t]$ is quantized using a simple uniform scalar quantizer with ${\bar B}_{\ell}$ bits quantization levels. 
 The transmitted signal of the $n$th antenna at RRH $\ell$ after applying the quantization is given by
\begin{align}
    {\bar x}_{\ell}^n[t] &= {x}_{\ell}^n[t] + {v}_{\ell}^n[t],
    \label{eq:Quantization_Error_Tx}
\end{align}
where ${v}_{\ell}^n[t]$ is the quantization noise of ${x}_{\ell}^n[t]$ which is assumed to be the complex Gaussian with zero-mean and variance $\mathbb{E}[|{v}_{\ell}^{n}[t]|^2 ]={\sigma}^2_{{v}_{\ell}^{n}}$, i.e., ${v}_{\ell}^{n}[t]\sim\mathcal{CN}\left({ 0},{\sigma}^2_{{v}_{\ell}^{n}}\right)$. From \cite{mezghani2007modified,zhang2016mixed,gersho2012vector}, the quantization noise variance when using the ${\bar B}_{\ell}$ bits uniform scalar quantizer with ${\bar B}_{\ell}\ge 6$ is tightly approximated as 
\begin{align}
 {\sigma}^2_{{v}_{\ell}^{n}} \simeq  \eta_{\ell}({\bar B}_{\ell})\cdot\mathbb{E}\left[|x_{\ell}^n[t]|^2\right]
	= \eta_{\ell}({\bar B}_{\ell})\cdot\left(\sum_{k=1}^{K} |f_{\ell,k}^n|^2P\right),\label{eq:quantizaton_noise_CU}
\end{align}
where $\eta_{\ell}({\bar B}_{\ell})=\frac{\pi\sqrt{3}}{2}2^{-2B_{\ell}}$. Therefore, the quantized signals $\left\{{\bar x}_{\ell}^1[t],\ldots, {\bar x}_{\ell}^N[t]\right\}$, each with $2{\bar B}_{\ell}$ bits, are reliably delivered from BBU to the $\ell$th RRH with the rate of 
\begin{align}
	\sum_{n=1}^NI({\bar x}_{\ell}^n[t];{x}_{\ell}^n[t])=\sum_{n=1}^N\log\left(1+\frac{\sum_{k=1}^{K} |f_{\ell,k}^n|^2P}{{\sigma}^2_{{v}_{\ell}^{n}}}\right)\simeq N\log\left(1+\eta_{\ell}({\bar B}_{\ell})^{-1}\right). \label{eq:backhaul_capacity_constraint_CU}
\end{align}
Using the rate expression in \eqref{eq:backhaul_capacity_constraint_CU}, BBU selects the number of quantization bits $B_{\ell}$ to minimize ${\sigma}^2_{{v}_{\ell}^{n}}$ while ensuring the fronthaul capacity constraint such that
\begin{align}
	{\bar B}_{\ell}^{\star}=\left\lfloor\frac{1}{2}\log_2\left(\frac{\pi\sqrt{3}}{2}\left(2^{\frac{C_{\ell}}{N}}-1\right)\right)\right\rfloor. \label{eq:optimal_B_CU}
\end{align}
It is remarkable that the quantization bits ${\bar B}_{\ell}^{\star}$ derived in \eqref{eq:optimal_B_CU} allows us to satisfy the fronthaul capacity constraints regardless of precoding strategies because it alters the quantization levels as a function of the norm of precoding vectors to meet the constraint. From the relationship between ${\sigma}^2_{{v}_{\ell}^{n}}$ and ${\bar B}_{\ell}$ in \eqref{eq:quantizaton_noise_CU}, the effective quantization noise variance ${\sigma}^2_{{v}_{\ell}^{n}}$ reduces by designing the precoding vectors $\sum_{k=1}^{K} |f_{\ell,k}^n|^2$ for each $n\in [N]$ and $\ell\in\mathcal{L}$ with a small norm. Therefore, our precoding strategy aims at minimizing the norm of precoding vectors $\sum_{k=1}^{K} |f_{\ell,k}^n|^2$ for each $n\in [N]$ and $\ell\in\mathcal{L}$. To explicitly represent ${\sigma}^2_{{v}_{\ell}^{n}}$ as a function of precoding vectors,  we define  a precoding matrix for RRH $\ell\in\mathcal{L}$ by ${\bf F}_{\ell}=[{\bf f}_{\ell,1},\hdots,{\bf f}_{\ell, K}]$. Then, the covariance matrix for the quantization noise in a compact form is 
\begin{align}
{\bf V}_{\ell} ({\bf F}_{\ell},{\bar B}_{\ell}^{\star})
	= \mathbb{E}[{\bf v}_{\ell}[t]{\bf v}_{\ell}[t]^{\sf H}]=P\cdot\eta_{\ell}^{\star}({\bar B}_{\ell}^{\star})\cdot{\sf diag}\left(\sum_{k=1}^{K}|f_{\ell,k}^1|^2,\hdots,\sum_{k=1}^{K}|f_{\ell,k}^N|^2\right),\label{eq:quantizaton_noise2}
\end{align}
where $\eta_{\ell}^{\star}({\bar B}_{\ell}^{\star})=\frac{\pi\sqrt{3}}{2}2^{-2{\bar B}_{\ell}^{\star}}$ and ${\bf v}_{\ell}[t]=\left[v_{\ell}^1[t],\ldots, v_{\ell}^N[t]\right]^{\sf T}$.

\vspace{0.3cm}
{\bf Ergodic spectral efficiency:} The received signal of the $k$th user is 
\begin{align}
  y_{k}[t] &= \sum_{\ell=1}^{L}{\bf h}_{\ell,k}^{\sf H}{\bf\bar x}_{\ell}[t] + z_k[t]\nonumber\\
    &= \sum_{\ell=1}^{L}{\bf h}_{\ell,k}^{\sf H}{\bf f}_{\ell,k}s_k[t]+\sum_{\ell=1}^{L}\sum_{i\ne k}{\bf h}_{\ell,k}^{\sf H}{\bf f}_{\ell,i}s_i[t]
  + \sum_{\ell=1}^{L}{{\bf h}_{\ell,k}^{\sf H}{\bf v}_{\ell}}[t]+ z_k[t],
  \label{eq: received_signal}
\end{align}
where $z_k[t]$ is the noise signal of the $k$th user, which is distributed as $\mathcal{CN}(0,\sigma^2)$. 
Then, the signal-to-interference-plus-noise ration (SINR) of the $k$th user is defined as
\begin{align}
  {\sf SINR}_k  
  &= \frac{\left|\sum_{\ell=1}^{L}{\bf h}_{\ell,k}^{\sf H}{\bf f}_{\ell,k}\right|^2}{\sum_{i\ne k}\left|\sum_{\ell=1}^{L}{\bf h}_{\ell,k}^{\sf H}{\bf f}_{\ell,i}\right|^2+  \sum_{\ell=1}^{L}{\bf h}_{\ell,k}^{\sf H}{\bf V}_{\ell} ({\bf F}_{\ell},{\bar B}_{\ell}^{\star})
{\bf h}_{\ell,k}/P +\sigma^2 /P}.
\end{align}
With noisy-and-incomplete CSIT, ${\bf \bar H}=\{{\bf \bar h}_{\ell,k}\}_{\ell\in\mathcal{L},k\in\mathcal{K}_{\ell}}$, the BBU estimates the instantaneous spectral efficiency of the $k$th downlink user, i.e., 
\begin{align}
    R_k\left({\bf\bar H}\right)=
    \mathbb{E}_{{\bf H}|{\bf\bar H}}\left[\log_2\left(1+{\sf SINR}_k\right)|{\bf\bar H}\right], \quad k\in\mathcal{K}_{\ell},
    \label{eq:Estimated_SE}
\end{align}
where the expectation is taken over both channel estimation and quantization errors. Therefore, by taking the expectation over every fading states, we obtain the ergodic spectral efficiency
\begin{align}
	{\bar R}_k = \left(1-\frac{\tau_u+\tau_d}{\tau_c}\right)\mathbb{E}_{{\bf\bar H}}\left[R_k\left({\bf\bar H}\right)\right]
	=\left(1-\frac{\tau_u+\tau_d}{\tau_c}\right)\mathbb{E}\left[\log_2\left(1+{\sf SINR}_k\right)\right], 
\end{align}
where $\tau_u$ and $\tau_d$ denote  the uplink and downlink channel training lengths respectively.

\vspace{0.5cm}

\section{Sum-Spectral Efficiency Maximization Problem}
In this section, we shall formulate a sum-spectral efficiency maximization problem with noisy-and-incomplete CSIT under a sparsely active RRH constraint.  To accomplish this, we first derive a lower bound of the instantaneous spectral efficiency. Then, we formulate the spare precoding optimization problem that maximizes the obtained lower bound of the instantaneous spectral efficiency under the sparsely active RRH constraint. 


\subsection{A Lower Bound of Instantaneous Spectral Efficiency}
 We begin by rewriting the received signal in \eqref{eq: received_signal} in terms of the noisy-and-quantized CSIT at BBU, i.e., ${\bf h}_{\ell,k}={\bf \hat h}_{\ell,k}+{\bf e}_{\ell,k}={\bf \bar h}_{\ell,k}+{\bf e}_{\ell,k}+{\bf q}_{\ell,k}$, which yields
\begin{align}
  y_{k}[t] &= 
  \sum_{\ell=1}^{L}\left[\left({\bf \bar h}_{\ell,k}+{\bf e}_{\ell,k}+{\bf q}_{\ell,k} \right)^{\sf H}\left(
  {\bf f}_{\ell,k}s_k[t]
  +\sum_{i\ne k}{\bf f}_{\ell,i}s_i[t] 
  +{\bf v}_{\ell}[t]
  \right)\right]+z_k[t]\nonumber\\
  &=\sum_{\ell=1}^{L}{\bf \bar h}_{\ell,k}^{\sf H}{\bf f}_{\ell,k}s_k[t]
  +\sum_{\ell=1}^{L}\sum_{i\ne k}{\bf \bar h}_{\ell,k}^{\sf H}{\bf f}_{\ell,i}s_i[t]+{\tilde z}_k[t],
  \label{eq: received_signal_2}
\end{align}
where ${\tilde z}_k[t]$ is the effective noise term, i.e., 
\begin{align}
    {\tilde z}_k[t]=\sum_{\ell=1}^{L}\sum_{i=1}^K \left({\bf e}_{\ell,k}+{\bf q}_{\ell,k}\right)^{\sf H}{\bf f}_{\ell,i}s_i[t]
    +\sum_{\ell=1}^{L}\left({\bf \bar h}_{\ell,k}+{\bf e}_{\ell,k}+{\bf q}_{\ell,k}\right)^{\sf H}{\bf v}_{\ell}[t]+z_k[t]. \label{eq:effective_noise}
\end{align}
Unfortunately, the effective noise $  {\tilde z}_k[t]$ is non-Gaussian because the product of two Gaussian random variables $s_i[t]$ and ${\bf e}_{\ell,k}+{\bf q}_{\ell,k}$ is not Gaussian. Harnessing the generalized mutual information  \cite{yoo2006capacity,medard2000effect,lapidoth2002fading,ding2010maximum}, in which the non-Gaussian noise is simply modeled as the Gaussian noise with a proper moment matching, we characterize a lower bound of the instantaneous spectral efficiency \cite{choi2018joint,choi2019joint}.  To accomplish this, we need to compute the variance of the effective noise  ${\tilde z}_k[t]$. Since $\mathbb{E}[ {\tilde z}_k[t]]=0$, the effective noise variance is
\begin{align}
	{\tilde \sigma}^2_{k}&=\mathbb{E}[ |{\tilde z}_k[t]|^2]\nonumber\\
	&= P\sum_{\ell=1}^{L}\sum_{i=1}^K  {\bf f}_{\ell,i}^{\sf H}\mathbb{E}\left[ {\bf e}_{\ell,k}{\bf e}_{\ell,k}^{\sf H}+{\bf q}_{\ell,k}{\bf q}_{\ell,k}^{\sf H}+{\bf e}_{\ell,k}{\bf q}_{\ell,k}^{\sf H}+{\bf q}_{\ell,k}{\bf e}_{\ell,k}^{\sf H}\right]{\bf f}_{\ell,i}
	+\sum_{\ell=1}^{L} \mathbb{E}\left[{\bf h}_{\ell,k}^{\sf H}{\bf v}_{\ell}[t]{\bf v}_{\ell}[t]^{\sf H}{\bf h}_{\ell,k}\right] +\sigma^2\\  
	&=P\sum_{\ell=1}^{L}\sum_{i=1}^K  {\bf f}_{\ell,i}^{\sf H}\left({\bf \Phi}_{\ell,k}+{\bf Q}_{\ell,k}(B_{\ell}^{\star})\right){\bf f}_{\ell,i} +\sum_{\ell=1}^L {\sf Tr}\left({\bf R}_{\ell,k}{\bf V}_{\ell} ({\bf F}_{\ell},{\bar B}_{\ell}^{\star})\right) +\sigma^2, \label{eq:var_eff_noise}
\end{align}
where the last equality follows from the fact that the channel estimation error noise is independent of the channel quantization noise, i.e., $\mathbb{E}\left[ {\bf e}_{\ell,k}{\bf q}_{\ell,k}^{\sf H}\right]={\bf 0}$. Invoking this effective noise variance, a lower bound of instantaneous spectral efficiency when using noisy-and-incomplete CSIT is
\begin{align}
    \sum_{k=1}^KR_k^{\sf low}=   \sum_{k=1}^K\log_2\left(1+\frac{\left|\sum_{\ell=1}^L{\bf\bar h}_{\ell,k}^{\sf H}{\bf f}_{\ell,k}\right|^2}{\sum_{i\ne k}\left|\sum_{\ell=1}^L{\bf \bar h}_{\ell,k}^{\sf H}{\bf f}_{\ell,i}\right|^2+{\tilde\sigma}_k^2/P}\right).\label{eq:lower_bound}
\end{align}
The sum-spectral efficiency in \eqref{eq:lower_bound} is the estimate of the instantaneous sum-spectral efficiency with limited channel knowledge, which will be used to find a joint solution for user clustering, beamforming, and power allocation in the sequel.

\subsection{Sparsely Active RRH Constraint }
 Let $S$ be the maximum number of active RRHs per the joint transmission, and it is assumed to be smaller than a total number of RRHs $L$ in the network, i.e., $S\le L$.  We also define an index set of active RRHs as 
\begin{align}
	\mathcal{A}=\{\ell: \|{\bf x}_{\ell}[t]\|_2^2\neq 0\}\subset \mathcal{L}.
\end{align}
It is true that $\sum_{k=1}^{K}\|{\bf f}_{\ell,k}\|_2^2 > 0$ if $\|{\bf x}_{\ell}[t]\|_2^2\neq 0$. Using this relation, to perform sparse-JT, we need to design the precoding vectors to satisfy the following group-sparsity condition: 
\begin{align}
	\sum_{\ell=1}^{L} {\bf 1}_{ \left\{\sum_{k=1}^{K}\|{\bf f}_{\ell,k}\|_2^2 > 0\right\}}\leq S, \label{eq:GS_constraint}
\end{align}
where ${\bf 1}_{{\mathcal{C}}}$ is an indicator function such that ${\bf 1}_{{\mathcal{C}}}=1$ if an event ${\mathcal{C}}$ is true and ${\bf 1}_{{\mathcal{C}}}=0$  otherwise. Our optimization task is to identify precoding vectors, $\{{\bf f}_{\ell,k}\}_{\ell\in\mathcal{L},k\in\mathcal{K}}$, to maximize the lower bound of the instantaneous spectral efficiency \eqref{eq:lower_bound} under the group-sparsity constraints \eqref{eq:GS_constraint}. This optimization problem is formulated as

\begin{subequations}
\label{eq:optimization_problem}
\begin{align}
    \mathscr{P}^1:~~~~&{\underset{\{{\bf f}_{\ell,k}\}_{\ell\in\mathcal{L},k\in\mathcal{K}}}{\text{arg~max}}}~ \sum_{k=1}^{K}\log_2\left(1+\frac{\left|\sum_{\ell=1}^{L}{\bf\bar h}_{\ell,k}^{\sf H}{\bf f}_{\ell,k}\right|^2}{\sum_{i\ne k}\left|\sum_{\ell=1}^{L}{\bf \bar h}_{\ell,k}^{\sf H}{\bf f}_{\ell,i}\right|^2+{\tilde\sigma}_k^2/P}\right)\label{eq:Objective_Function}  ,\\
    &\text{subject to} ~~ \sum_{k=1}^{K}\|{\bf f}_{\ell,k}\|_2^2\le (1+{\eta}_{\ell}^{\star}({\bar B}_{\ell}^{\star}))^{-1},~~~\forall\ell \in \mathcal{L}, \label{eq: sum-power_RRH}\\
    &~~~~~~~~~~~~~\sum_{\ell=1}^{L} {\bf 1}_{ \left\{\sum_{k=1}^{K}\|{\bf f}_{\ell,k}\|_2^2 > 0\right\}}\leq S,\label{eq: sparsity_Constraint}
\end{align}
\end{subequations}

where the $L$ inequalities in \eqref{eq: sum-power_RRH} correspond to the per-RRH power constraint, $\mathbb{E}[\|{\bf\bar x}_{\ell}\|_2^2]\le P$. Obtaining the global optimal solution for this optimization problem even without a group-sparsity constraint is highly non-trivial, because the objective function is non-convex with respective to precoding vectors. Additionally, the group-sparsity constraint makes the problem a combinatorial optimization. 




%
 
 \subsection{Reformation to a Generalized Sparse-PCA Problem}
We explain how the optimization problem \eqref{eq:optimization_problem} can be reformulated in a generalized sparse-PCA problem. The following proposition elucidates the connection between them.

\prop{Let ${\bf f}\in\mathbb{C}^{LNK\times 1}$ be a network-wide precoding vector by concatenating all precoding vectors to form a large-dimensional optimization variable, namely,
\begin{align}
     {\bf{f}} &= [\underbrace{{\bf{f}}_{1,1}^{\sf{H}} ,\cdots, {\bf{f}}_{L,1}^{\sf{H}}}_{{\bf f}_1^{\sf H}} ,\cdots, \underbrace{{\bf {f}}_{1,k}^{\sf {H}} ,\cdots, {\bf {f}}_{L,k}^{\sf {H}}}_{{\bf f}_{k}^{\sf H}},\cdots, \underbrace{{\bf {f}}_{1,K}^{\sf {H}} ,\cdots, {\bf {f}}_{L,K}^{\sf {H}}}_{{\bf f}_K^{\sf H}}]^{\sf {H}}\in \mathbb{C}^{LNK\times 1}. \label{eq:network-wide_BFV}
\end{align} 
We also define large-dimensional positive semidefinite matrices ${\bf A}_k\in\mathbb{C}^{LNK\times LNK}$ and ${\bf B}_k\in\mathbb{C}^{LNK\times LNK}$ such that ${\bf f}^{\sf H}{\bf A}_k{\bf f}$ and ${\bf f}^{\sf H}{\bf B}_k{\bf f}$ are the total received power and the interference power received at user $k$th, which are \begin{align}
{\bf  A}_{k}&={\bf I}_K \otimes \left({\bf \bar h}_k{\bf \bar h}_k^{\sf H}+{\bf \Phi}_k+{\bf Q}_k\left(B_{\ell}^{\star}\right)\right) + \frac{\sum_{\ell=1}^L {\sf Tr}\left({\bf R}_{\ell,k}{\bf V}_{\ell} ({\bf F}_{\ell},{\bar B}_{\ell}^{\star})\right) +\sigma^2}{P\cdot \sum_{\ell=1}^L(1+{\eta}_{\ell}^{\star}({\bar B}_{\ell}^{\star}))^{-1}} {\bf I}_{LNK}\in\mathbb{C}^{LNK\times LNK},\label{eq:A}\\
{\bf  B}_{k}&={\bf  A}_{k}-{\bf a}_k{\bf a}_k^{\sf T}\otimes {\bf \bar h}_k{\bf \bar h}_k^{\sf H}\in\mathbb{C}^{LNK\times LNK}, \label{eq:B}
\end{align}
where ${\bf\bar h}_{k} = \left[{{\bf \bar  h}}_{1,k}^{\sf H}, \hdots, {{\bf \bar h}}_{L,k}^{\sf H}\right]^{\sf H}$, ${\bf\Phi}_k={\sf diag}\left({\bf\Phi}_{1,k},\hdots,{\bf\Phi}_{L,k}\right)$, and ${\bf Q}_k(B_{\ell}^{\star})={\sf diag}({\bf Q}_{1,k}(B_{\ell}^{\star}),\hdots,{\bf Q}_{L,k}(B_{\ell}^{\star}))$. Then, the optimization problem $\mathscr{P}^1$ is equivalent to the following problem:
\begin{subequations}
\begin{align}
    \mathscr{P}^2:~~~~&{\underset{{\bf f}\in \mathbb{C}^{LNK\times 1}}{\text{arg~max}}}~\log_2\left(\prod_{k=1}^{K}\frac{{{\bf f}^{\sf H}{\bf A}_{k}{\bf f}}}{{{\bf f}^{\sf H}{\bf  B}_{k}{\bf f}}}\right),\\
    &\text{subject to} \sum_{\ell=1}^{L} {\bf 1}_{ \left\{\sum_{k=1}^K\|{\bf f}_{\ell,k}\|_2^2 > 0\right\}}\leq S.
\end{align}\label{eq:S-PCA_optimization_problem}
\end{subequations}}
\begin{IEEEproof}
The key idea is that we reformulate the objective function \eqref{eq:Objective_Function} to a product of the Rayleigh quotients by representing the optimization variables in a high dimensional space as in \cite{choi2018joint,choi2019joint,han2020distributed}. Specifically, using the aggregated vectors, ${\bf \bar h}_{k}$ and ${\bf f}_{k}$, we arrange the received signal representation of the $k$th user in \eqref{eq: received_signal} to a compact form as
\begin{align}
  y_{k}[t]
  = {\bf \bar h}_k^{\sf H}{\bf f}_ks_k[t] 
  + \sum_{i\ne k}^K{\bf \bar h}_k^{\sf H}{\bf f}_is_i[t] 
  + {\tilde z}_k[t],
  \label{eq: received_signal_vec}
\end{align}
where the effective noise $ {\tilde z}_k[t] $ is defined with the aggregated channel estimation and quantization error vectors ${\bf  e}_{k} = \left[{{\bf e}}_{1,k}^{\sf H} ,\ldots, {{\bf e}}_{L,k}^{\sf H}\right]^{\sf H} $, ${\bf  q}_{k} = \left[{{\bf q}}_{1,k}^{\sf H} ,\ldots, {{\bf q}}_{L,k}^{\sf H}\right]^{\sf H} $ and ${\bf v}[t] = \left[{\bf v}_{1}[t]^{\sf H} ,\ldots, {\bf v}_{L}[t]^{\sf H}\right]^{\sf H}$ as
\begin{align}
    {\tilde z}_k[t] = \sum_{i=1}^K \left({\bf e}_{k}-{\bf q}_{k}\right)^{\sf H}{\bf f}_{i}s_i[t]
    + {\bf h}_{k}^{\sf H} {\bf v}[t]+z_k[t],\label{eq:simp_noise}
\end{align}
Then, the variance of effective noise can be rewritten with respective to the aggregate precoding vectors as
\begin{align}
    {\tilde \sigma}_k^2=\mathbb{E}\left[|{\tilde z}_k[t]^2|\right]=P\sum_{i=1}^{K}{\bf f}_{i}^{\sf H}\left({\bf \Phi}_k+{\bf Q}_k\left(B_{\ell}^{\star}\right)\right){\bf f}_{i}+\sum_{\ell=1}^L {\sf Tr}\left({\bf R}_{\ell,k}{\bf V}_{\ell} ({\bf F}_{\ell},{\bar B}_{\ell}^{\star})\right) +\sigma^2.
    \label{eq:effective_noise_vec}
\end{align}
Note that ${\bf e}_{\ell,k}$, ${\bf q}_{\ell,k}$, ${\bf \Phi}_{\ell,k}$, and ${\bf Q}_{\ell,k}(B_{\ell}^{\star})$ becomes zero vectors and matrices when $\ell\notin\mathcal{L}_{k}$. Furthermore, we relax the per-RRH power constraint, $\sum_{k=1}^{K}\|{\bf f}_{\ell,k}\|_2^2\le (1+{\eta}_{\ell}^{\star}({\bar B}_{\ell}^{\star}))^{-1}$ for all $\ell \in \mathcal{L}$, to the network-wide sum-power constraint, i.e., $\sum_{\ell=1}^{L}\sum_{k=1}^{K}\|{\bf f}_{\ell,k}\|_2^2=\|{\bf f}\|_2^2= \sum_{\ell=1}^L(1+{\eta}_{\ell}^{\star}({\bar B}_{\ell}^{\star}))^{-1}$. This relaxation reduces $L$ equality constraints to a single equality constraint. Then, harnessing the large-dimensional network-wide precoding vector ${\bf f}$, our objective function in \eqref{eq:Objective_Function} is written in a form of the product of Rayleigh quotients as
\begin{align}
\sum_{k=1}^{K}\log_2\left(1+\frac{\left|\sum_{\ell=1}^{L}{\bf\bar h}_{\ell,k}^{\sf H}{\bf f}_{\ell,k}\right|^2}{\sum_{i\ne k}\left|\sum_{\ell=1}^{L}{\bf \bar h}_{\ell,k}^{\sf H}{\bf f}_{\ell,i}\right|^2+{\tilde\sigma}_k^2/P}\right)
&=\sum_{k=1}^{K}\log_2\left(\frac{\sum_{i=1}^{K}\left|\sum_{\ell = 1}^{L}{\bf\bar h}_{\ell,k}^{\sf H}{\bf f}_{\ell,i}\right|^2+{\tilde\sigma}_k^2/P}{\sum_{i\ne k}^{K}\left|\sum_{\ell=1}^L{\bf \bar h}_{\ell,k}^{\sf H}{\bf f}_{\ell,i}\right|^2+{\tilde\sigma}_k^2/P}\right)\nonumber\\
&=\log_2\left(\prod_{k=1}^{K}\frac{{{\bf f}^{\sf H}{\bf A}_{k}{\bf f}}}{{{\bf f}^{\sf H}{\bf  B}_{k}{\bf f}}}\right).\label{eq:Objective_Rayleigh}
\end{align} 
Since the objective function \eqref{eq:Objective_Rayleigh} is invariant to scale of any real value $\alpha \in \mathbb{R}$ on ${\bf f}$, i.e,. $
	\log_2\left(\prod_{k=1}^{K}\frac{{{\bf f}^{\sf H}{\bf A}_{k}{\bf f}}}{{{\bf f}^{\sf H}{\bf  B}_{k}{\bf f}}}\right)=	\log_2\left(\prod_{k=1}^{K}\frac{{\alpha{\bf f}^{\sf H}{\bf A}_{k}\alpha{\bf f}}}{{\alpha{\bf f}^{\sf H}{\bf  B}_{k}\alpha{\bf f}}}\right)$, we discard the sum-power constraint to further simplify the optimization problem.  Therefore, the sum-spectral efficiency maximization problem in \eqref{eq:optimization_problem} is equivalent to \eqref{eq:S-PCA_optimization_problem}, which completes the proof.
\end{IEEEproof}

This reformulated optimization problem is interesting because it can be interpreted with a lens through a generalized sparse-PCA problem in machine learning \cite{zou2006sparse}.  To shed further light on the significance of the reformulation in \eqref{eq:S-PCA_optimization_problem}, we will provide a more detailed explanation at the end of this section. 

\subsection{Tractable Relaxation for Group-Sparsity Constraint}

Unfortunately, the reformulated optimization problem \eqref{eq:S-PCA_optimization_problem} in Proposition 1 is still a non-convex and combinatorial optimization problem. In this section, we take a non-convex approximation to relax the group-sparsity constraint in a tractable quadratic form. 

\prop{Let ${\bf C}_{\ell}\in\mathbb{C}^{LNK\times LNK}$ be a positive semidefinite matrix with a block diagonal structure defined as ${\bf C}_{\ell} = {\bf I}_{K} \otimes {\bf \tilde C}_{\ell},$ where ${\bf\tilde C}_{\ell} = {\bf a}_{\ell}{\bf a}_{\ell}^{\sf T}\otimes\epsilon^{-1}{\bf I}_N+\frac{1}{L}{\bf I}_{NL}$. Using this matrix, the approximation for the group-sparsity constraint has a form of the product of Rayleigh quotients, i.e.,  
\begin{align}
    \sum_{\ell=1}^{L} {\bf 1}_{ \left\{\sum_{k=1}^{K}\|{\bf f}_{\ell,k}\|_2^2 > 0\right\}}\approx \log_2\prod_{\ell=1}^L\left({\bf f}^{\sf H}{\bf  C}_{\ell}{\bf f}\right)^{\mu_{\epsilon}}. \label{eq:apprx_sparse_constraint}
\end{align}
Then, the optimization problem $\mathscr{P}^2$ with the approximate constraint boils down to the following optimization problem:
\begin{subequations}
\label{eq:S-PCA_optimization_problem_approx}
\begin{align}
    \mathscr{P}^3:~~~~&{\underset{{\bf f}\in \mathbb{C}^{LNK\times 1}}{\text{arg~max}}}~\log_2\left(\prod_{k=1}^{K}\frac{{{\bf f}^{\sf H}{\bf A}_{k}{\bf f}}}{{{\bf f}^{\sf H}{\bf  B}_{k}{\bf f}}}\right),\\
    &\text{subject to}~  \log_2\prod_{\ell=1}^L\left({\bf f}^{\sf H}{\bf  C}_{\ell}{\bf f}\right)^{\mu_{\epsilon}}\leq S.
\end{align}
\end{subequations}
}
\begin{IEEEproof}
From \cite{sriperumbudur2011majorization}, the indicator function for event set $\{|x|>0\}$ is \begin{align}
   {\bf 1}_{\{|x|>0\}} = \lim_{\epsilon\to0}{\frac{\log_2(1+|x|/\epsilon)}{\log_2(1+1/\epsilon)}}.\label{eq:Indicator_Apprx}
\end{align}
Using this limiting value, for sufficiently small $\epsilon>0$, it is possible to make a tight approximation for the group-sparsity constraint in a quadratic form with respective to the precoding vectors:
\begin{align}
\sum_{\ell=1}^{L} {\bf 1}_{ \left\{\sum_{k=1}^{K}\|{\bf f}_{\ell,k}\|_2^2 > 0\right\}}
&\approx  \log_2\left[\prod_{\ell=1}^{L}{\left(1+\epsilon^{-1}\left(\sum_{k=1}^{K}{\bf f}_{\ell,k}^{\sf H}{\bf f}_{\ell,k} \right)\right)^{\mu_{\epsilon}}} \right],\nonumber\\
&=\log_2\prod_{\ell=1}^L\left({\bf f}^{\sf H}{\bf  C}_{\ell}{\bf f}\right)^{\mu_{\epsilon}},
\end{align}
where $\mu_{\epsilon} = 1/{\log_2{(1+\epsilon^{-1})}}$. 
 With this non-convex relaxation, our optimization problem \eqref{eq:S-PCA_optimization_problem} simplifies as \eqref{eq:S-PCA_optimization_problem_approx}.  
\end{IEEEproof}

Notice that the relaxed group-sparsity constraint is still a non-convex function with respective to ${\bf f}$. Nevertheless, this relaxation is a tractable form for our optimization framework, which will be explained in the next section.

\subsection{ Interpretation}
We provide a detailed explanation to clearly provide the motivation for the reformulation of a sparse-PCA form.  Suppose a single-user case, i.e., $K=1$. In this case, finding the sparse precoding vector ${\bf f} \in \mathbb{C}^{LN\times 1}$ to maximize the sum-spectral efficiency under a block sparsity constraint as in \eqref{eq:S-PCA_optimization_problem} can be reformulated as a well-known sparse-PCA problem:
\begin{subequations}\label{eq:sparse_PCA_L0_1.4}
\begin{align}
    &{\underset{{\bf f} \in \mathbb{C}^{LN\times 1}}{\text{arg~max}}}~{{\bf f}^{\sf H}\left({\bf A}_{1}+{(\sigma^2P)^{-1}}{\bf I}\right){\bf f}},\label{eq:sparse_PCA_L0_1.4_1}\\
    &\text{subject to } \|{\bf f}\|_2^2=P,\label{eq:sparse_PCA_L0_1.4_2}\\
    &~~~~~~~~~~~~~\!\|{\bf \tilde f}\|_0\leq S,\label{eq:sparse_PCA_L0_1.4_3}
\end{align}
\end{subequations}
where ${\bf \tilde f} = \left[\|{\bf f}_{1,1}\|_2^2,\cdots,\|{\bf f}_{\ell,1}\|_2^2,\cdots,\|{\bf f}_{L,1}\|_2^2\right]^{\sf T}\in\mathbb{R}^{L\times 1}$. The optimal precoding solution ${\bf f}^{\star}$ for  \eqref{eq:sparse_PCA_L0_1.4} is a principal eigenvector of ${\bf A}_1+(\sigma^2P)^{-1}{\bf I}$ with the block sparsity constraint $S$. Since it is a NP-hard problem \cite{luo2008dynamic, yu2013multicell, hong2013joint}, there is no algorithm to find the optimal solution with a polynomial time complexity.  To overcome this challenge, the use of the $\ell_1$ norm convex relaxation method, which provides a convex lower bound of the $\ell_0$ norm function, has been used, which reformulates the problem as a convex optimization problem:
\begin{subequations}\label{eq:sparse_PCA_L1_1.4}
\begin{align}
    &{\underset{{\bf f}\in \mathbb{C}^{LN\times 1}}{\text{arg~max}}}~{{\bf f}^{\sf H}\left({\bf A}_{1}+{(\sigma^2P)^{-1}}{\bf I}\right){\bf f}},\label{eq:sparse_PCA_L1_1.4_1}\\
    &\text{subject to } \|{\bf f}\|_2^2=P,\label{eq:sparse_PCA_L1_1.4_2}\\
    &~~~~~~~~~~~~~\!\|{\bf \tilde f}\|_1\leq S+\delta,\label{eq:sparse_PCA_L1_1.4_3}
\end{align}
\end{subequations}
for some $\delta>0$.  Thanks to the convexity, the problem \eqref{eq:sparse_PCA_L1_1.4} can be solved by applying the algorithm introduced in \cite{journee2010generalized} with computational complexity $\mathcal{O}(JL^2N^2)$. 

We generalize this single-user case to a multi-user case $K\neq 1$. Then, our sparse joint precoding design problem becomes \begin{subequations}
\label{eq:sparse_PCA_L0_multi_1.4}
\begin{align}
    &{\underset{{\bf f} \in \mathbb{C}^{LNK\times 1}}{\text{arg~max}}}~\log_2\prod_{k=1}^{K}\frac{{{\bf f}^{\sf H}{\bf A}_{k}{\bf f}}}{{{\bf f}^{\sf H}{\bf  B}_{k}{\bf f}}},\label{eq:sparse_PCA_L0_multi_1.4_1}\\
    &\text{subject to } \|{\bf f}\|_2^2=P,\label{eq:sparse_PCA_L0_multi_1.4_2}\\
    &~~~~~~~~~~~~~\!\|{\bf \bar f}\|_0\leq S,\label{eq:sparse_PCA_L0_multi_1.4_3}
\end{align}
\end{subequations}
where ${\bf \bar f} = \left[\|{\bf\bar f}_{1}\|_2^2,\cdots,\|{\bf\bar f}_{\ell}\|_2^2,\cdots,\|{\bf\bar f}_{L}\|_2^2\right]^{\sf T}\in\mathbb{R}^{L\times 1}$ and ${\bf \bar f}_{\ell} = \left[{\bf f}_{\ell,1}^{\sf H},\cdots,{\bf f}_{\ell,k}^{\sf H},\cdots,{\bf f}_{\ell,K}\right]^{\sf H}\in\mathbb{C}^{NK\times 1}$. Contrast to the single-user case, our optimization task is to find a common block-sparse principal vector ${\bf f}$ that simultaneously maximizes $\frac{ {\bf f}^{\sf H}{\bf A}_{k}{\bf f}}{ {\bf f}^{\sf H}{\bf  B}_{k}{\bf f}}$ for $k\in [K]$.  When interpreting the identification of sparse vector ${\bf f}$ to maximize $\frac{ {\bf f}^{\sf H}{\bf A}_{k}{\bf f}}{ {\bf f}^{\sf H}{\bf  B}_{k}{\bf f}}$ as a task that finds a sparse linear classifier separating two classes in linear discriminant analysis (LDA), this problem can be interpreted as a generalized sparse LDA problem in a multi-task setting. Unfortunately,  finding such ${\bf f}$ is non-trivial even using the convex relaxation technique in \eqref{eq:sparse_PCA_L1_1.4}, because the optimization problem still remains non-convex. Therefore, instead of applying the $\ell_1$ norm convex relaxation method, we approximate the sparsity constraint in \eqref{eq:sparse_PCA_L0_multi_1.4_3} into a log-product form, which is a tractable non-convex function for our optimization framework, as $\|{\bf \bar f}\|_0= \sum_{\ell=1}^{L} {\bf 1}_{ \left\{\sum_{k=1}^{K}\|{\bf f}_{\ell,k}\|_2^2 > 0\right\}}\approx \log_2\prod_{\ell=1}^L\left({\bf f}^{\sf H}{\bf  C}_{\ell}{\bf f}\right)^{\mu_{\epsilon}}.$ Using this non-convex approximation method, we finally arrive at \eqref{eq:S-PCA_optimization_problem_approx}.

\section{Local Optimality Conditions}
This section devotes to derive local optimality conditions for the relaxed sum-spectral efficiency maximization problem \eqref{eq:S-PCA_optimization_problem_approx}. The following theorems show the first- and the second-order necessary conditions for a local optimal solution.




\begin{theorem}
\label{thm1}
{\bf(The first-order necessary condition)} Let $\gamma({\bf f},\lambda)=\frac{\prod_{k=1}^{K}{\bf f}^{\sf H}{\bf A}_{k}{\bf f}}{\prod_{k=1}^{K}{\bf f}^{\sf H}{\bf  B}_{k}{\bf f}\prod_{\ell=1}^L\left({\bf f}^{\sf H}{\bf  C}_{\ell}{\bf f}\right)^{\mu_{\epsilon}\lambda}}$. Any stationary point ${\bf f}\in \mathbb{C}^{LNK\times 1}$ for problem \eqref{eq:S-PCA_optimization_problem_approx} is an eigenvector of the following functional generalized eigenvalue problem:
\begin{align}
    {\bf \bar A}\left({\bf f}\right){\bf f}
    =\gamma\left({\bf f},\lambda\right){\bf \bar B}\left({\bf f},\lambda\right){\bf f},
    \label{eq:cond1}
\end{align}
where
\begin{gather*}
    {\bf \bar A}\left({\bf f}\right) = \left(\prod_{k=1}^{K}{\bf f}^{\sf H}{\bf A}_{k}{\bf f}\right)\sum_{i=1}^{K}\frac{{\bf A}_i}{{\bf f}^{\sf H}{\bf A}_i{\bf f}},\nonumber\\
    {\bf \bar B}\left({\bf f},\lambda\right) = \left(\prod_{k=1}^{K}{\bf f}^{\sf H}{\bf  B}_{k}{\bf f}\prod_{\ell=1}^L\left({\bf f}^{\sf H}{\bf  C}_{\ell}{\bf f}\right)^{\mu_{\epsilon}\lambda}\right)\left(\sum_{i=1}^{K}{\frac{{\bf B}_i}{{\bf f}^{\sf H}{\bf B}_i{\bf f}}}+ \sum_{i=1}^{L}{\frac{\mu_{\epsilon}\lambda{\bf C}_i}{{\bf f}^{\sf H}{\bf C}_i{\bf f}}}\right).
\end{gather*}
In addition, the Lagrange multiplier $\lambda$ is chosen so that ${\bf f}$ satisfies 
\begin{align}
	 \log_2\prod_{\ell=1}^L\left({\bf f}^{\sf H}{\bf  C}_{\ell}{\bf f}\right)^{\mu_{\epsilon}}= S.\label{eq:cond2}
\end{align}
\end{theorem}
\begin{IEEEproof}
See Appendix A.
\end{IEEEproof}
Theorem \ref{thm1} implies that one can find a stationary point of the non-convex optimization problem in \eqref{eq:S-PCA_optimization_problem_approx} by solving the functional generalized eigenvalue problem. In particular, the objective function normalized by the sparsity level, i.e., $\gamma({\bf f},\lambda)=\frac{\prod_{k=1}^{K}{\bf f}^{\sf H}{\bf A}_{k}{\bf f}}{\prod_{k=1}^{K}{\bf f}^{\sf H}{\bf  B}_{k}{\bf f}\prod_{\ell=1}^L\left({\bf f}^{\sf H}{\bf  C}_{\ell}{\bf f}\right)^{\mu_{\epsilon}\lambda}}$, can be interpreted as an eigenvalue for the functional generalized eigenvalue problem \eqref{eq:S-PCA_optimization_problem_approx}. Since both $\left[\sum_{i=1}^{K}\frac{{\bf A}_i}{{\bf f}^{\sf H}{\bf A}_i{\bf f}}\right]$ and $\left[\sum_{i=1}^{K}{\frac{{\bf B}_i}{{\bf f}^{\sf H}{\bf B}_i{\bf f}}}+ \sum_{i=1}^{L}{\frac{\mu_{\epsilon}\lambda{\bf C}_i}{{\bf f}^{\sf H}{\bf C}_i{\bf f}}}\right]$ matrices are full-rank with probability one, there are $LNK$ distinct eigenvectors, i.e., stationary points, each with distinct objective function value. This fact allows us to roughly visualize the global landscape of this non-convex function. Since we need to maximize $\gamma({\bf f},\lambda)$, we postulate that the eigenvector corresponding to the maximum eigenvalue can be globally optimal solution. Finding the maximum eigenvector, however, is a very challenging task over all possible ${\bf f}\in \mathbb{C}^{LNK\times 1}$ and $\lambda \in \mathbb{R}$. Instead, we find a local optimal solution that satisfies the first-order condition in Theorem \ref{thm1} and the following second-order condition.

\begin{theorem}
\label{thm2}
{\bf (The second-order necessary condition)} Let ${\bf f}^{\star}$ and $\lambda^{\star}$ be the solution Theorem \ref{thm1}. This stationary point ${\bf f}^{\star}$ is a local-optimal solution, provided that 
\begin{align}
    &\rho_{\sf min}\left(\sum_{i=1}^{K}   \frac{{\bf A}_i^{\sf H}{\bf f}^{\star}({\bf f}^{\star})^{\sf H}{\bf A}_i}{\left(({\bf f}^{\star})^{\sf H}{\bf A}_i{\bf f}^{\star}\right)^2} \right) >
    \rho_{\sf max}\left(\sum_{i=1}^{K}   \frac{{\bf B}_i^{\sf H}{\bf f}^{\star}({\bf f}^{\star})^{\sf H}{\bf B}_i}{\left(({\bf f}^{\star})^{\sf H}{\bf B}_i{\bf f}^{\star}\right)^2}      + \sum_{i=1}^{L}\mu_{\epsilon}\lambda^{\star}{\frac{{\bf C}_i^{\sf H}{\bf f}^{\star}({\bf f}^{\star})^{\sf H}{\bf C}_i}{ ( ({\bf f}^{\star})^{\sf H} {\bf C}_i{\bf f})^2}} \right).    \label{eq:Sub_Opt_Condition}
\end{align}
\end{theorem}
 \begin{IEEEproof}
See Appendix B.
\end{IEEEproof}
Theorem \ref{thm2} implies that to have a direction of strictly negative curvature at the saddle point ${\bf f}^{\star}$, it is sufficient that the minimum eigenvalue of $\sum_{i=1}^{K}   \frac{{\bf A}_i^{\sf H}{\bf f}^{\star}({\bf f}^{\star})^{\sf H}{\bf A}_i}{\left(({\bf f}^{\star})^{\sf H}{\bf A}_i{\bf f}^{\star}\right)^2}$ is greater than the maximum eigenvalue of $\sum_{i=1}^{K}   \frac{{\bf B}_i^{\sf H}{\bf f}^{\star}({\bf f}^{\star})^{\sf H}{\bf B}_i}{\left(({\bf f}^{\star})^{\sf H}{\bf B}_i{\bf f}^{\star}\right)^2}      + \sum_{i=1}^{L}\mu_{\epsilon}\lambda^{\star}{\frac{{\bf C}_i^{\sf H}{\bf f}^{\star}({\bf f}^{\star})^{\sf H}{\bf C}_i}{ ( ({\bf f}^{\star})^{\sf H} {\bf C}_i{\bf f})^2}}$. This allows us to check whether a saddle point ${\bf f}^{\star}$ is the local optimal solution for the non-convex optimization problem with the eigenvalue test. The maximum and the minimum eigenvalues can be computed using both power and inverse power iteration algorithms. 

 \section{Sparse Joint Transmission}
\begin{algorithm}[t]
\caption{SPARSE JOINT TRANSMISSION ALGORITHM.}
Initialization: $n=t=0,~ {\bf f}^{(0)}={\sf ZF},\lambda_+^{(0)},~ \lambda_-^{(0)}~ {\text{and }} \epsilon$\;
    \While{$\left|\log_2\prod_{\ell=1}^{L}\left(\left({\bf f}^{(t)}\right)^{\sf H}{\bf C}_{\ell}{\bf f}^{(t)}\right)^{\mu_{\epsilon}}-S\right|\ge\epsilon$}
    {
        $n\gets n+1$\;
        
        $\lambda^{(n)}\gets \frac{\lambda_+^{(n-1)}+\lambda_-^{(n-1)}}{2}$\;
        
        \While{$\|\gamma\left({\bf f}^{(t-1)}\right)-\gamma\left({\bf f}^{(t)}\right)\|_2\ge\epsilon$}
        {
            $t\gets t+1$\;
            ${\bf f}^{(t)} \gets \left[{\bf \bar B}\left({\bf f}^{(t-1)},\lambda^{(n)}\right)\right]^{-1}{\bf \bar A}\left({\bf f}^{(t-1)}\right){\bf f}^{(t-1)}$\;
            ${\bf f}^{(t)} \gets \frac{\sqrt{\sum_{\ell=1}^L(1+{\eta}_{\ell}^{\star}({\bar B}_{\ell}^{\star}))^{-1}}\cdot {\bf f}^{(t)}}{\max\left\{\left\{\left[\sum_{k=1}^{K}\|{\bf f}_{\ell,k}\|_2^2\right]^{1/2}\right\}_{\ell\in\mathcal{L}}\right\}}$\;
        }
        
        \uIf{${\sf sign}(g(\lambda_-^{(n-1)}))\ne{\sf sign}(g(\lambda^{(n)}))$}
        {$\lambda_+^{(n)}\gets\lambda^{(n)}$\\$\lambda_-^{(n)}\gets\lambda_-^{(n-1)}$}
        \Else{$\lambda_+^{(n)}\gets\lambda_+^{(n-1)}$\\$\lambda_-^{(n)}\gets\lambda^{(n)}$}
        
    }
\label{alg:GPI_AlG}
\end{algorithm}


 From Theorem \ref{thm1} and Theorem \ref{thm2}, we have established the local-optimality conditions for the network-wide precoding vector. To obtain such ${\bf f}^{\star}$, however, we need to solve a large-dimensional nonlinear system of equations.  As a result, it is essential to design an algorithm that finds the local-optimal solution in a computationally efficient manner. By generalizing the method in \cite{choi2018joint,choi2019joint,han2020distributed}, we propose a computationally efficient algorithm to find such a local-optimal solution. 

The proposed sparse-JT algorithm finds a sparse network-wide precoding vector in an iterative manner.  In the $t$th iteration, using the previously identified sparse precoding vector ${\bf f}^{(t-1)}$ and the Lagrange multiplier $\lambda^{(n)}$, we construct the functional matrices ${\bf \bar A}\left({\bf f}^{(t-1)}\right)$ and $ {\bf \bar B}\left({\bf f}^{(t-1)},\lambda^{(n)}\right)$. Then, using the generalized power iteration algorithm initially introduced in \cite{lee2008achievable}, we update the sparse precoding vector such that
\begin{align}
	  {\bf f}^{(t)}=:\left[ {\bf \bar B}\left({\bf f}^{(t-1)},\lambda^{(n)}\right) \right]^{-1}
	  {\bf \bar A}\left({\bf f}^{(t-1)}\right){\bf f}^{(t-1)},
\end{align}
with normalization $\frac{\sqrt{L}{\bf f}^{(t)}}{\|{\bf f}^{(t)}\|_2}$ until it converges on the first eigenvector within a sufficiently small positive value $\epsilon$, i.e., $\|{\bf f}^{(t-1)}-{\bf f}^{(t)}\|_2\leq \epsilon$. Using this convergent solution ${\bf f}^{(t)}$, the algorithm checks whether it satisfies the group-sparsity condition:
\begin{align}
	\left|\log_2\prod_{\ell=1}^L\left(({\bf f}^{(t)})^{\sf H}{\bf  C}_{\ell}{\bf f}^{(t)}\right)^{\mu_{\epsilon}}  - S\right| \leq \epsilon.
\end{align}
If the group-sparsity condition is satisfied, the algorithm moves to next step. Otherwise, Algorithm 1 updates $\lambda^{(n)}$ using the bisection method, which is a simple root-finding technique for any continuous monotonic function.  Applying this method,  Algorithm 1 repeatedly bisects an interval $\left[\lambda^{(n)}_{+}, \lambda^{(n)}_{-}\right]$ with the function values $g\left(\lambda^{(n)}_{+}\right)=\left(\log_2\prod_{\ell=1}^L\left(({\bf f}^{(t)})^{\sf H}{\bf  C}_{\ell}{\bf f}^{(t)}\right)^{\mu_{\epsilon}}  - S\right)>0$ and $g\left(\lambda^{(n)}_{-}\right)=\left(\log_2\prod_{\ell=1}^L\left(({\bf f}^{(t)})^{\sf H}{\bf  C}_{\ell}{\bf f}^{(t)}\right)^{\mu_{\epsilon}}  - S\right)<0$, where ${\bf f}^{(t)}= \left[ {\bf \bar B}\left({\bf f}^{(t-1)},\lambda^{(n-1)}\right) \right]^{-1}
{\bf \bar A}\left({\bf f}^{(t-1)}\right){\bf f}^{(t-1)}$. The iterations of the bisection method end when the function value approaches an interval $[-\epsilon, +\epsilon]$, where $\epsilon$ is a predetermined precision error constant. After finishing the iterations, in the last step, we apply project the sparse precoding vector onto the per-RRH power constraint sets to ensure the power constraint per RRH as  \begin{align}
\frac{(1+{\eta}_{\ell}^{\star}({\bar B}_{\ell}^{\star}))^{-1/2}\cdot{\bf f}^{(t)}}{\max\left\{\left\{\left[\sum_{k=1}^{K}\|{\bf f}_{\ell,k}\|_2^2\right]^{1/2}\right\}_{\ell\in\mathcal{L}}\right\}}.
\end{align}
The proposed algorithm is summarized in Algorithm \ref{alg:GPI_AlG}. 
 

 
{\bf Remark 1 (Validation for local optimality):} Using system-level simulations, we numerically validate the local optimality of the sparse-JT solution for the proposed algorithm. Interestingly, the solution ${\bf f}^{\star}$ obtained from Algorithm \ref{alg:GPI_AlG} satisfies the local optimality condition derived in Theorem \ref{thm2} in every fading realization when the algorithm starts with zero-forcing (ZF) precoding solution as the initial point. 

 
\begin {table}[t]
\caption {System Level Simulation Assumptions.} \vspace{-0.1cm}\label{tab:Sys_Assumption} 
  	 \begin{center}
  \begin{tabular}{ l  c }
    \hline\hline
    Parameters & Value  \\ \hline
        Topology of RRH & Densely deployed in [2000m $\times$ 2000m]  \\ 
        Topology of User & Randomly distributed in [2000m $\times$ 2000m]  \\ 
         Bandwidth & 10MHz \\
         Carrier frequency & 2GHz \\
         RRH transmission power & 40dBm \\
         Noise power & -113dB\\
         Spatial channel model & Spatially correlated model\\
         Path-loss model & Okumaura-Hata model\\
         RRH/UE height & 32m/1.5m\\
         Channel estimation & Imperfect \\ 
        Quantization bits ${B}^{\star}$ & 3$\sim$6 bits\\\hline
  \end{tabular}
\end{center}\vspace{-0.3cm}
\end {table}
%

{\bf  Remark 2  (Downlink data sharing overhead reduction):} Our sparse joint downlink transmission can reduce the downlink data sharing overhead because BBU sends downlink data symbols ${\bf x}_{\ell}[t]$ to the selected $S$ RRHs via a finite-rate fronthaul links, each with $C_{\ell}$ for $\ell\in \mathcal{A}$.	 Therefore, the downlink data sharing overhead diminishes as the number of active RRHs $S$ decreases.

{\bf  Remark 3 (CSI acquisition overhead reduction):} Our sparse joint transmission can also reduce the CSI sharing overhead. RRH $\ell\in \mathcal{L}$ estimates CSI for $U_{\ell}^{\star}<K$ users that provide the highest received power, and share them with BBU for the precoding construction. Then, the uplink overhead for the CSI acquisition can be diminished by the factor of $\frac{\sum_{\ell=1}^LU_{\ell}^{\star}}{KL}$.

{\bf  Remark 4 (Computational complexity reduction in precoding):} Our precoding algorithm requires the computational complexity order of $\mathcal{O}(JL^2N^2K)$, while the WMMSE-based sparse beamforming algorithm needs the computational complexity of $\mathcal{O}\left(J(KLN)^{3.5}\right)$. Therefore, our algorithm is much less complex as $K$, $L$, and $N$ increase.

\section{Simulations Results}
\begin{figure}[t]
	\centering
    \includegraphics[width=6cm]{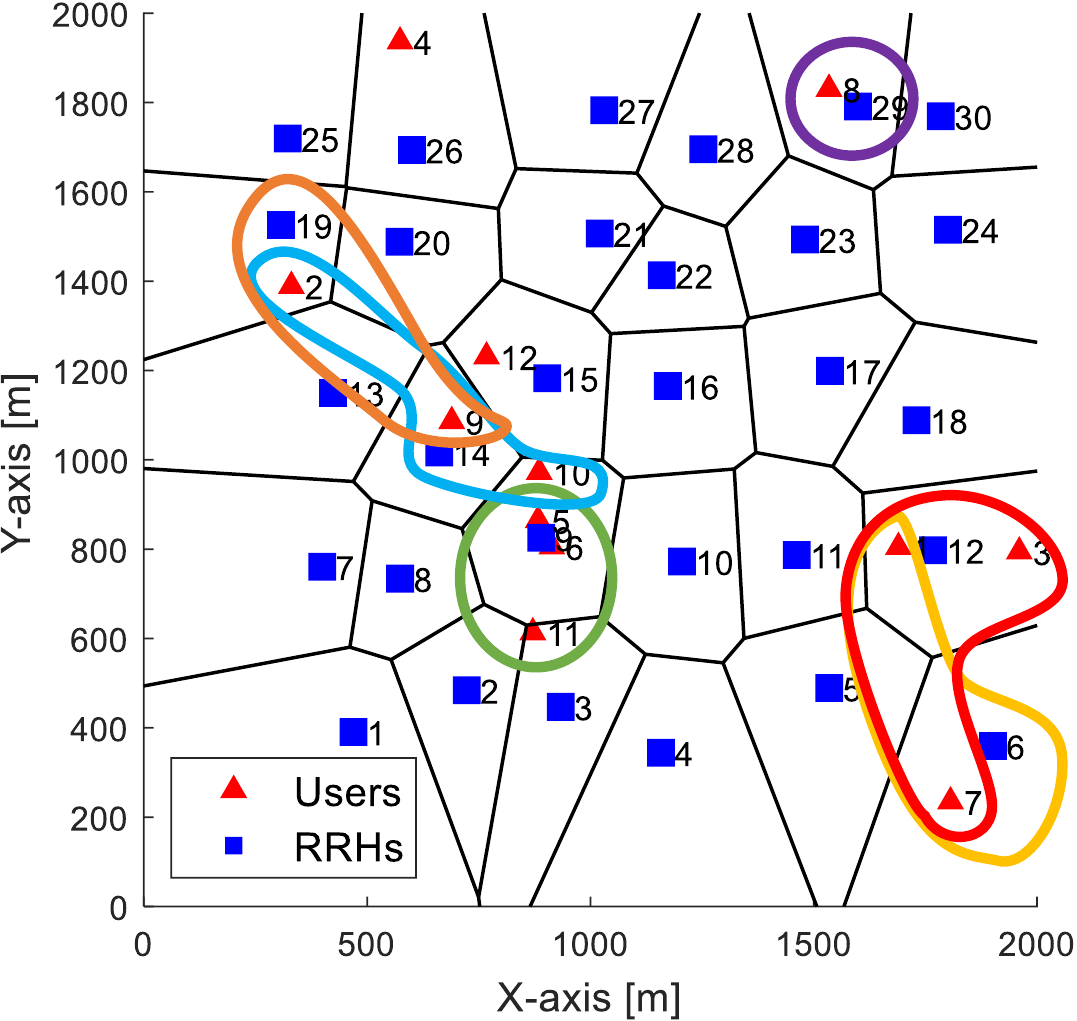}
  \caption{The illustration of a dynamic clustering under $(C_{\ell},U_{\ell}^{\star},B_{\ell}^{\star},{\bar B}_{\ell}^{\star})=(3\text{ }Gbps,6,6,12)$ and $S=6$.} \label{f_Dynamic_Clustering}
\end{figure}

In this section, we provide system-level simulation results to compare the performance of the proposed sparse-JT with those of the existing transmit precoding schemes. The topology and simulation parameters are summarized in Table \ref{tab:Sys_Assumption}. 

To understand the joint transmission solution ${\bf f}^{\star}$ obtained from our algorithm, we provide an illustrative example. Consider a simulation setting in which $(C_{\ell},U_{\ell}^{\star},B_{\ell}^{\star},{\bar B}_{\ell}^{\star})=(3\text{ Gbps}~,6,6,12)$ and $[L,K,S]=[30, 10, 6]$. In this case, as depicted in Fig. \ref{f_Dynamic_Clustering}, the solution ${\bf f}^{\star}$ obtained from our algorithm provides a set of active RRHs as $\mathcal{A}=\{6,9,12,14,19,29\}$. To be specific, RRH 9 and 29 perform single-cell MU and SU MIMO transmissions, while RRH 6, 12, 14, and 19 perform the joint transmission by partially sharing the transmit data streams. Here, the data sharing pattern for the joint transmission is depicted by different group colors. For instance, RRH 12 serves three users in cell 6 and 12.  Meanwhile, RRH 6 supports two users in the cells. In addition, the users in cells 15 and 26 are discarded because the number of active RRHs $S$ can be less than $K$. 

To fairly compare with the proposed method, we consider the following existing RRH clustering and precoding methods:
\begin{itemize}
	\item RRH-centric clustering with zero-forcing beamforming (RCC-ZF):  This scheme first selects a set of active RRHs with the $S$ most significant aggregated channel gains from all users. Then, the conventional ZF precoding is applied using the selected RRH set.
    \item Sparse-JT clustering with zero-forcing beamforming (SC-ZF):  In this method, a set of active RRHs and corresponding serving users are chosen from the proposed sparse-JT precoding method. Specifically, from the obtained solution ${\bf f}^{\star}$ of the proposed sparse-JT, the SC-ZF searches activated ${\bf f}_{\ell,k}$ for all $\ell\in\mathcal{L}$ and $k\in\mathcal{K}_{\ell}$. Then, the conventional ZF precoding is applied using the selected cluster set. 
    \item WMMSE \cite{christensen2008weighted}: We also consider the WMMSE based sparse precoding method. 
    After obtaining the precoding vector, it choose the best $S$ precoding vector in terms of power allocation and the other precoding vectors are set to be zero. 
\end{itemize}


\begin{figure}[t]
	\centering
    \includegraphics[width=0.45\linewidth]{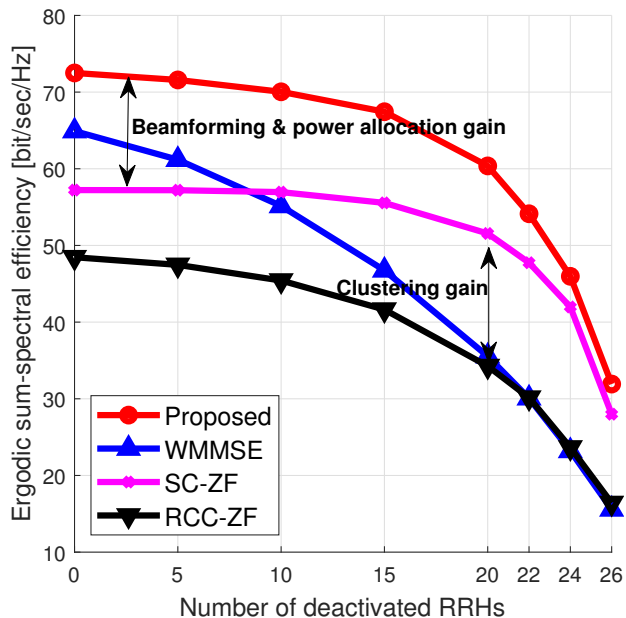}
  \caption{The ergodic sum-spectral efficiency comparison when decreasing the number of cooperative RRHs. We consider the system configuration of $(L,N,K)=(30,4,12)$ under noisy-and-incomplete CSIT with $(C_{\ell},U_{\ell}^{\star},B_{\ell}^{\star},{\bar B}_{\ell}^{\star})=(3,6,6,12)$.} \label{Fig_sim_1}
\end{figure}


{\bf{Trade-off between ergodic sum-spectral efficiency and the number of active RRHs: }}
Fig. \ref{Fig_sim_1} shows how the ergodic sum-spectral efficiency alters when the number of cooperative RRHs decreases. To elucidate the effect of the number of cooperative active RRHs for JT under the limited fronthaul capacity, we assume that the fronthaul capacity is $C_{\ell}=3$ Gbps and choose the $(U_{\ell}^{\star},B_{\ell}^{\star},{\bar B}_{\ell}^{\star})=(6,6,12)$.  As can be seen in Fig. \ref{Fig_sim_1}, the proposed sparse-JT method provides significant gains compared to all existing methods regardless of the number of active RRHs. In particular, the proposed sparse-JT achieves a better trade-off performance (i.e., less performance degradation) than all other existing JT strategies when increasing the number of deactivated RRHs. One interesting observation from Fig. \ref{Fig_sim_1} is that, on the one hand, the beamforming and the power allocation strategies are more crucial than the RRHs clustering strategy because the inter-RRH interference is a major bottleneck when all RRHs are active. On the other hand,  the clustering strategy becomes more significant than the beamforming and power allocation as the number of active RRHs decreases. 


%

{\bf{Effects of noisy-and-incomplete CSIT: }}
\begin{figure}[t]
     \centering
    \includegraphics[width=0.45\linewidth]{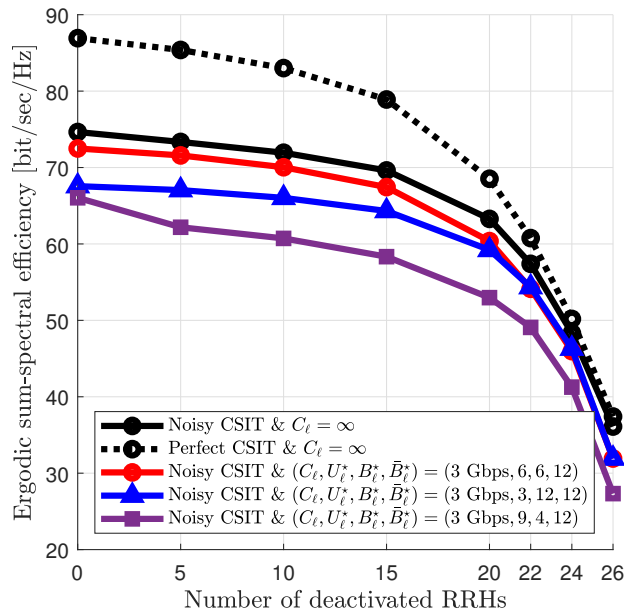}
  \caption{The ergodic sum-spectral efficiency comparison as the number of cooperative RRHs decreases under diverse CSIT assumptions; perfect CSIT, noisy CSIT, and noisy-and-incomplete CSIT.} \label{Fig_sim_3}
\end{figure}
Fig. \ref{Fig_sim_3} shows the effects of noisy-and-incomplete CSIT on the ergodic sum-spectral efficiency. To see the effects on the channel estimation errors, we consider the perfect CSIT case as a benchmark (a black dotted line). As can be seen in Fig. \ref{Fig_sim_3}, the ergodic sum-spectral efficiency obtained in the noisy CSIT (a black solid line) shows the performance degradation compared to the perfect CSIT case, but the performance loss decreases as the activated RRHs become sparse. Furthermore, to gauge the effects on both the quantization noise and incomplete channel knowledge under the finite-rate fronthaul capacity, i.e., $C_{\ell}=3$ Gbps, we consider three different channel compression parameter sets; $(U_{\ell}^{\star},B_{\ell}^{\star},{\bar B}_{\ell}^{\star})=(6,6,12),\text{ }(9,4,12),\text{ and }(12,3,12)$. Fig. \ref{Fig_sim_3} shows that the ergodic sum-spectral efficiency additionally degrades by both quantization noise and incomplete channel knowledge. When $(U_{\ell}^{\star},B_{\ell}^{\star},{\bar B}_{\ell}^{\star})=(6,6,12)$, the performance loss caused by quantization noise and absent channel knowledge becomes negligible. In contrast, we observe a severe performance degradation when the compression parameters are chosen as $(U_{\ell}^{\star},B_{\ell}^{\star},{\bar B}_{\ell}^{\star})=(9,4,12)$ and $(12,3,12)$, respectively. To interpret these results, we provide numerical results in the next subsection.


{\bf{Effects of CSIT compression strategies: }}
\begin{figure}[t]
\subfigure[]{
\includegraphics[width=0.455\linewidth]{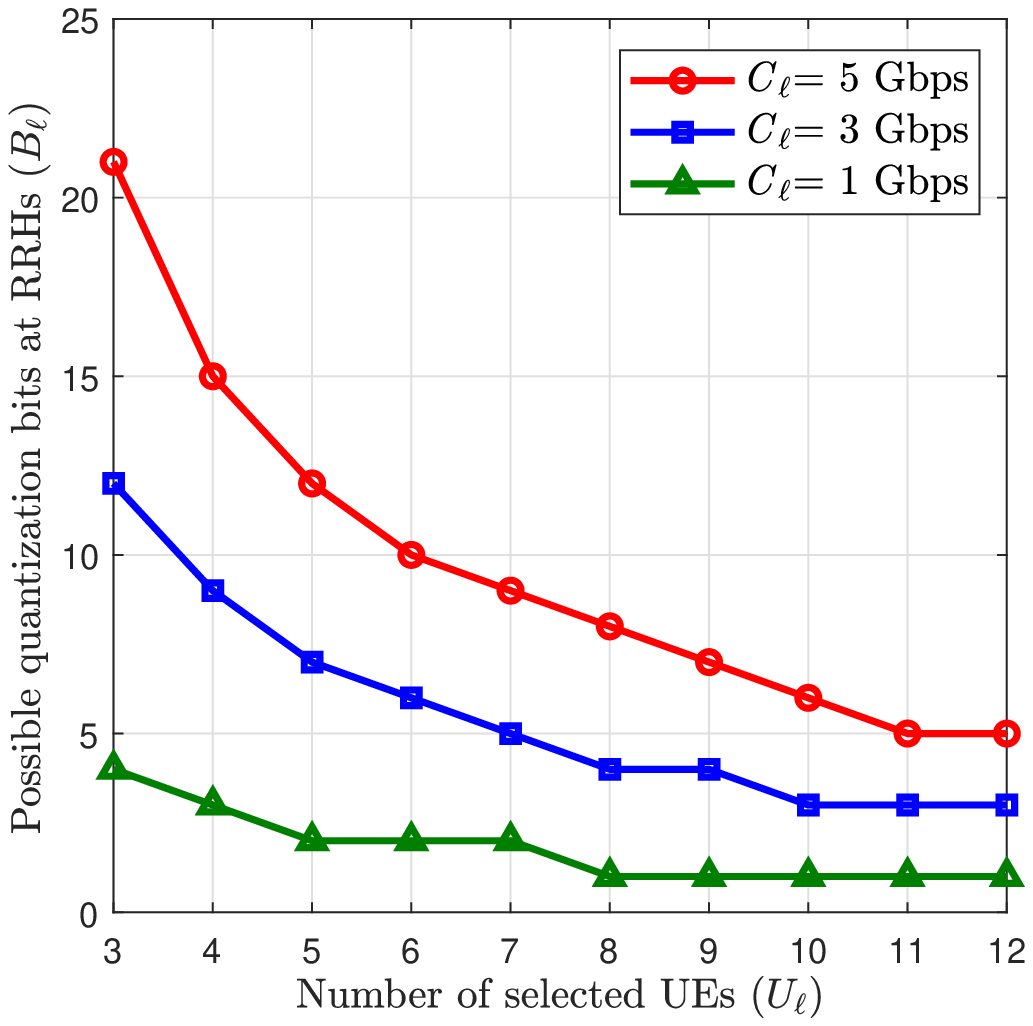}%
\label{Fig_sim_3_2}
}
\hspace*{\fill}
\subfigure[]{
\includegraphics[width=0.455\linewidth]{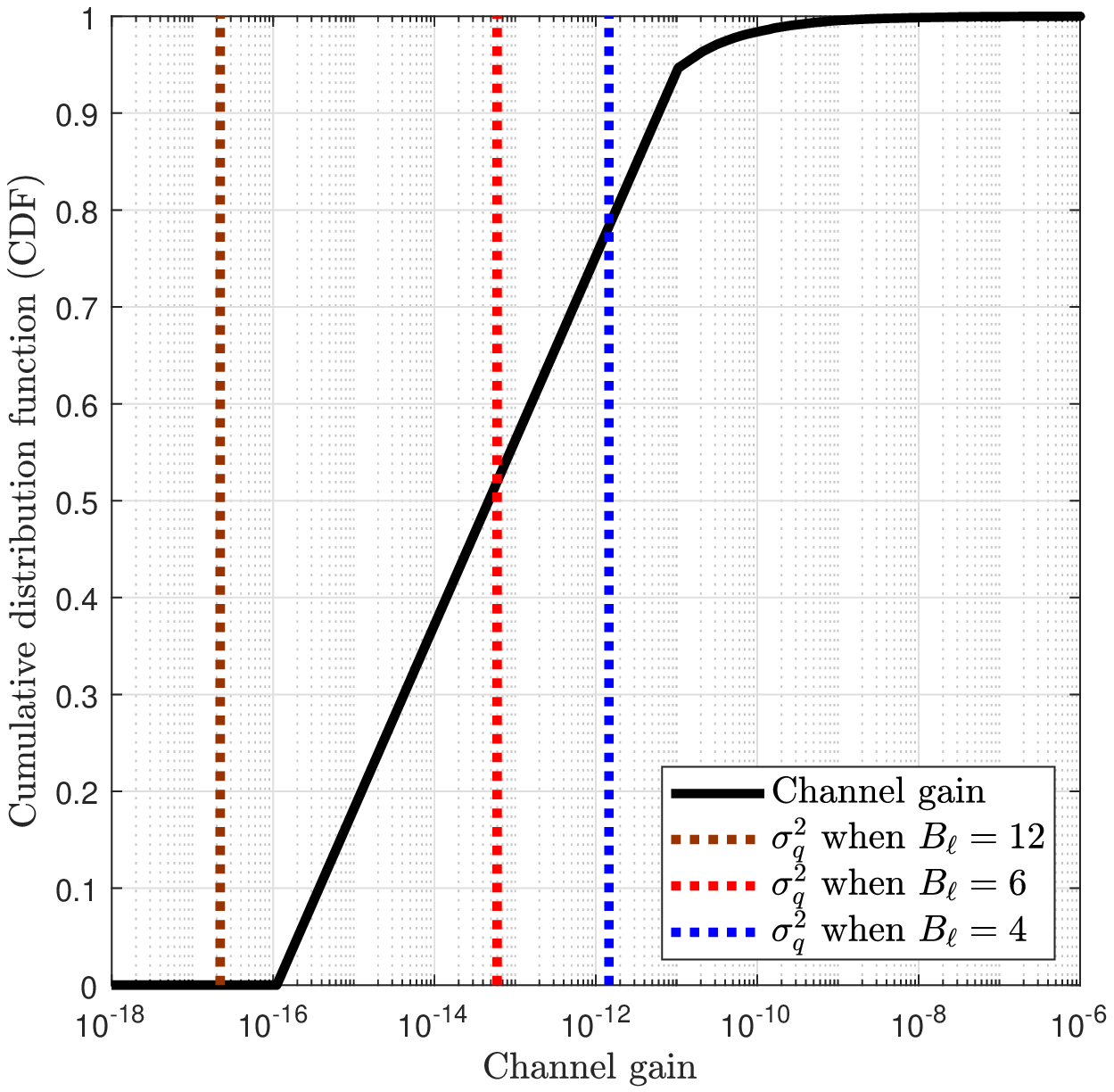}%
\label{Fig_sim_3_3}
}
\caption{(a) The possible quantization bits $B_{\ell}$ in a specific fronthaul capacity $C_{\ell}$ and the number of selected users $U_{\ell}$ and (b) the cumulative distribution function of channel gain and channel quantization noise levels.
}
\label{Fig_sim_5}
\end{figure}
Fig. \ref{Fig_sim_5}-(a) illustrates a trade-off between $B_{\ell}$ and $U_{\ell}$ when the fronthaul capacity has finite rates, $C_{\ell}\in\{1,3,5\}$ Gbps. As can be seen in Fig. \ref{Fig_sim_5}-(a), the quantization bits $B_{\ell}$ is inversely proportional to the number of selected users. To suitably choose the CSI compression parameters, $(U_{\ell}^{\star},B_{\ell}^{\star})$, we need to understand the channel gain distribution compared to the quantization noise level according to different $B_{\ell}$, which is depicted in Fig. \ref{Fig_sim_5}-(b). When $B_{\ell}=6$, the probability that the channel gain is greater than the quantization noise level, $\sigma_q^2=5.9869e^{-14}$, is approximately 0.5, which implies that $U_{\ell}=0.5K=6$ users should be selected to quantize their channels. However, when $B_{\ell}=4$, the probability that the channel gain is greater than the quantization noise level, $\sigma_q^2=1.4519e^{-12}$, is about 0.1. This means that only $U_{\ell}=\lfloor0.2K\rfloor=3$ user should be selected for the quantization. Nevertheless, when $B_{\ell}=4$, the nine users are selected under $C_{\ell}=3$ Gbps as shown in Fig. \ref{Fig_sim_5}-(a). Therefore, the CSI of the most users disappears by the high quantization noise level except for the CSI of three users. In other words, for a fixed $C_{\ell}$, we need to carefully select both $U_{\ell}$ and $B_{\ell}$ so that the CSI are efficiently delivered to the BBU.

{\bf{Effects of limited CSI acquisition: }}
\begin{figure}[t]
     \centering
    \includegraphics[width=0.45\linewidth]{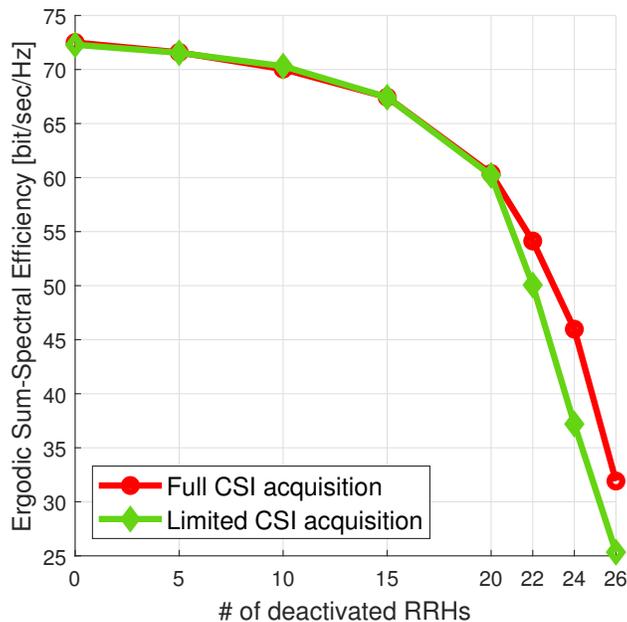}
  \caption{The ergodic sum-spectral efficiency comparison under different CSI acquisition assumptions when $(C_{\ell},U_{\ell}^{\star},B_{\ell}^{\star},{\bar B}_{\ell}^{\star})=(3,6,6,12)$.} \label{Fig_sim_8}
\end{figure}
To elucidate the effects of the limited CSI acquisition on the ergodic sum-spectral efficiency performance, we provide a numerical comparison with the full CSI acquisition case. As shown in Fig. \ref{Fig_sim_8}, there is no performance loss when the number of activated RRHs is sufficiently larger than the number of users, i.e., $S>K$. However, as the activated RRHs are sparse compared with the total number of users $K$, the method harnessing full CSI acquisition achieves a higher ergodic spectral efficiency. This result implies that more CSI overhead is required when activating RRHs sparse to increase the spectral efficiency in C-RANs.

\section{Conclusion}
This paper presented a novel sparse joint transmission method for a scalable C-RAN with noisy-and-incomplete CSIT and limited fronthaul capacity. The proposed sparse-JT aimed at maximizing a lower bound of the sum-spectral efficiency by jointly identifying a set of cooperative RRHs, precoding for beamforming and compression, and power control solutions. To accomplish this, a novel tractable non-convex optimization problem for the sum-spectral efficiency maximization was introduced under sparsely active RRH constraints. To solve the optimization problem, the sparse-JT algorithm that guarantees to identify a local-optimal solution was proposed. Simulation results demonstrated that the proposed sparse-JT offers significant gains over the existing joint transmission techniques in terms of the ergodic sum-spectral efficiency regardless of all system parameters.

One promising future research direction would investigate the sparse-JT with a user-centric clustering
method and the benefits of using deep-learning in the design of the sparse-JT.

 \appendix
 
 \subsection{Proof for Theorem 1}
 \proof

We commence by defining the Lagrange function:
\begin{align}
	    {\mathcal{L}}({\bf f},\lambda) &= \log_2\left(\prod_{k=1}^{K}\frac{{{\bf f}^{\sf H}{\bf A}_{k}{\bf f}}}{{{\bf f}^{\sf H}{\bf  B}_{k}{\bf f}}}\right)-\lambda \left(\log_2\prod_{\ell=1}^L\left({\bf f}^{\sf H}{\bf  C}_{\ell}{\bf f}\right)^{\mu_{\epsilon}}-S\right)\nonumber\\
	    &=\log_2\left(\frac{\prod_{k=1}^{K}{\bf f}^{\sf H}{\bf A}_{k}{\bf f}}{\prod_{k=1}^{K}{\bf f}^{\sf H}{\bf  B}_{k}{\bf f}\prod_{\ell=1}^L\left({\bf f}^{\sf H}{\bf  C}_{\ell}{\bf f}\right)^{\mu_{\epsilon}\lambda}}\right) +\lambda S, \label{eq:Lagrangian}
\end{align}
where $\lambda$ is the Lagrange multiplier. To find a stationary point, we take the partial derivatives of $ {\mathcal{L}}({\bf f},\lambda)$ with respective to ${\bf f}$ and $\lambda$ and set to them zero.  Let $f({\bf f})=\prod_{k=1}^{K}{\bf f}^{\sf H}{\bf A}_{k}{\bf f}$, $g({\bf f})=\prod_{k=1}^{K}{\bf f}^{\sf H}{\bf  B}_{k}{\bf f}$, and $
    h({\bf f},\lambda)=\prod_{\ell=1}^L\left({\bf f}^{\sf H}{\bf  C}_{\ell}{\bf f}\right)^{\mu_{\epsilon}\lambda}$. Then, the first condition $\nabla_{{\bf f}^{\sf H}}\mathcal{L({\bf f},\lambda)}=0$ is an equivalent one to solve $\nabla_{{\bf f}^{\sf H}}\gamma({\bf f},\lambda)=0$ by discarding the invariant constant $\lambda S$.
\begin{align}
    &\nabla_{{\bf f}^{\sf H}}\gamma({\bf f},\lambda)=0\nonumber\\
    &\Leftrightarrow\gamma({\bf f},\lambda)\left\{\sum_{i=1}^{K}{\frac{{\bf A}_i{\bf f}}{{\bf f}^{\sf H}{\bf A}_i{\bf f}}}-\sum_{i=1}^{K}{\frac{{\bf B}_i{\bf f}}{{\bf f}^{\sf H}{\bf B}_i{\bf f}}}-\sum_{i=1}^{L}{\frac{\mu_\epsilon\lambda{\bf C}_i{\bf f}}{{\bf f}^{\sf H}{\bf C}_i{\bf f}}}\right\}=0.
    \label{eq:proof_1}
\end{align}
Rearranging the condition \eqref{eq:proof_1}, we obtain 
\begin{align}
    {\bf \bar A}\left({\bf f}\right){\bf f}
    =\gamma\left({\bf f},\lambda\right){\bf \bar B}\left({\bf f},\lambda\right){\bf f}.
\end{align}
 
We also take the partial derivatives of $ {\mathcal{L}}({\bf f},\lambda)$ with respective to $\lambda$ and set to them zero.
\begin{align}
    \nabla_{\lambda}{\mathcal{L}}({\bf f},\lambda)=\log_2\prod_{\ell=1}^L\left({\bf f}^{\sf H}{\bf  C}_{\ell}{\bf f}\right)^{\mu_{\epsilon}}-S=0.\label{eq:condition_2}
\end{align}
The condition in \eqref{eq:condition_2} simplifies to 
\begin{align}
    \log_2\prod_{\ell=1}^L\left({\bf f}^{\sf H}{\bf  C}_{\ell}{\bf f}\right)^{\mu_{\epsilon}}=S.
\end{align}
This completes the proof.
\endproof

\subsection{Proof for Theorem 2}
\proof

To prove the local-optimality claim, it is sufficient to show that the extended Hessian matrix considering constraint sets at a stationary point is negative definite. To accomplish this, we first derive the extended Hessian matrix evaluated at an arbitrary point ${\bf f}\in \mathbb{C}^{LNK\times 1}$, which is given by
\begin{align}
     \nabla_{{\bf f}^{\sf H}}^2\gamma({\bf f},\lambda)
     &=2\left\{\nabla_{{\bf f}^{\sf H}}\gamma({\bf f},\lambda)\right\}\left(
     \sum_{i=1}^{K}\frac{{\bf A}_i{\bf f}}{{\bf f}^{\sf H}{\bf A}_i{\bf f}}
     -\sum_{i=1}^{K}\frac{{\bf B}_i{\bf f}}{{\bf f}^{\sf H}{\bf B}_i{\bf f}}
     -\sum_{i=1}^{L}\frac{\mu_{\epsilon}\lambda{\bf C}_i{\bf f}}{{\bf f}^{\sf H}{\bf C}_i{\bf f}}\right)^{\sf H}\nonumber\\
     &~~~~~~~~~~~~~~~~~~~~+2\gamma({\bf f},\lambda)\left\{\nabla_{{\bf f}^{\sf H}}\left(
     \sum_{i=1}^{K}\frac{{\bf A}_i{\bf f}}{{\bf f}^{\sf H}{\bf A}_i{\bf f}}
     -\sum_{i=1}^{K}\frac{{\bf B}_i{\bf f}}{{\bf f}^{\sf H}{\bf B}_i{\bf f}}
     -\sum_{i=1}^{L}\frac{\mu_{\epsilon}\lambda{\bf C}_i{\bf f}}{{\bf f}^{\sf H}{\bf C}_i{\bf f}}\right)\right\}.\label{eq:Hessian_matrix_1}
\end{align}
By plugging a stationary point ${\bf f}^{\star}$ and ${\lambda}^{\star}$obtained from Theorem 1 into \eqref{eq:Hessian_matrix_1}, it follows that
\begin{align}
     \nabla_{{\bf f}^{\sf H}}^2\gamma({\bf f}^{\star},\lambda^{\star})&=
     2\gamma({\bf f}^{\star},\lambda^{\star})
     \left\{
     \sum_{i=1}^{K}\frac{{\bf A}_i}{\left(({\bf f}^{\star})^{\sf H}{\bf A}_i{\bf f}^{\star}\right)}-\sum_{i=1}^{K}\frac{{\bf B}_i}{\left(({\bf f}^{\star})^{\sf H}{\bf B}_i{\bf f}^{\star}\right)}-\sum_{i=1}^{L}\frac{\mu_{\epsilon}\lambda^{\star}{\bf C}_i}{\left(({\bf f}^{\star})^{\sf H}{\bf C}_i{\bf f}^{\star}\right)}\right\}\nonumber\\
     &+2\gamma({\bf f}^{\star},\lambda^{\star})\left\{
     \sum_{i=1}^{K}\frac{-2{\bf A}_i{\bf f}^{\star}\left({\bf f}^{\star}\right)^{\sf H}{\bf A}_i}{(\left({\bf f}^{\star}\right)^{\sf H}{\bf A}_i{\bf f}^{\star})^2}
     +\sum_{i=1}^{K}\frac{2{\bf B}_i{\bf f}^{\star}\left({\bf f}^{\star}\right)^{\sf H}{\bf B}_i}{(\left({\bf f}^{\star}\right)^{\sf H}{\bf B}_i{\bf f}^{\star})^2}
     +\sum_{i=1}^{K}\frac{2\mu_{\epsilon}\lambda^{\star}{\bf C}_i{\bf f}^{\star}\left({\bf f}^{\star}\right)^{\sf H}{\bf C}_i}{(\left({\bf f}^{\star}\right)^{\sf H}{\bf C}_i{\bf f}^{\star})^2}
     \right\}.
     \label{eq:Hessian_matrix_2}
\end{align}
In \eqref{eq:Hessian_matrix_2}, the terms in first line $2\gamma({\bf f}^{\star},\lambda)
     \left\{
     \sum_{i=1}^{K}\frac{{\bf A}_i}{\left(({\bf f}^{\star})^{\sf H}{\bf A}_i{\bf f}^{\star}\right)}\!-\!\sum_{i=1}^{K}\frac{{\bf B}_i}{\left(({\bf f}^{\star})^{\sf H}{\bf B}_i{\bf f}^{\star}\right)}\!-\!\sum_{i=1}^{L}\frac{\mu_{\epsilon}\lambda^{\star}{\bf C}_i}{\left(({\bf f}^{\star})^{\sf H}{\bf C}_i{\bf f}^{\star}\right)}\right\}$ become zero from the result of Theorem 1. As a result, the extended Hessian matrix simplifies to
     \begin{align}
        \nabla_{{\bf f}^{\sf H}}^2\gamma({\bf f}^{\star},\lambda^{\star})
        = 4\gamma({\bf f}^{\star},\lambda^{\star})\left\{
     \sum_{i=1}^{K}\frac{{\bf B}_i{\bf f}^{\star}\left({\bf f}^{\star}\right)^{\sf H}{\bf B}_i}{(\left({\bf f}^{\star}\right)^{\sf H}{\bf B}_i{\bf f}^{\star})^2}
     +\sum_{i=1}^{K}\frac{\mu_{\epsilon}\lambda^{\star}{\bf C}_i{\bf f}^{\star}\left({\bf f}^{\star}\right)^{\sf H}{\bf C}_i}{(\left({\bf f}^{\star}\right)^{\sf H}{\bf C}_i{\bf f}^{\star})^2}
     -\sum_{i=1}^{K}\frac{{\bf A}_i{\bf f}^{\star}\left({\bf f}^{\star}\right)^{\sf H}{\bf A}_i}{(\left({\bf f}^{\star}\right)^{\sf H}{\bf A}_i{\bf f}^{\star})^2}
     \right\}
        .\label{eq:Hessian_matrix_3}
     \end{align}
     In \eqref{eq:Hessian_matrix_3}, the first term $\gamma({\bf f}^{\star},\lambda^{\star})$ is a positive scalar value and all the remaining terms are the summation of positive-definite matrices due to the fact that $A_i$, $B_i$, and $C_i$ are Hermitian  matrices for all $i\in\{1,\hdots,K\}$. It means that if the minimum eigenvalue of $\sum_{i=1}^{K}\frac{{\bf A}_i{\bf f}^{\star}\left({\bf f}^{\star}\right)^{\sf H}{\bf A}_i}{(\left({\bf f}^{\star}\right)^{\sf H}{\bf A}_i{\bf f}^{\star})^2}$ is bigger than the maximum eigenvalue of the $\sum_{i=1}^{K}\frac{{\bf B}_i{\bf f}^{\star}\left({\bf f}^{\star}\right)^{\sf H}{\bf B}_i}{(\left({\bf f}^{\star}\right)^{\sf H}{\bf B}_i{\bf f}^{\star})^2}
     +\sum_{i=1}^{K}\frac{\mu_{\epsilon}\lambda^{\star}{\bf C}_i{\bf f}^{\star}\left({\bf f}^{\star}\right)^{\sf H}{\bf C}_i}{(\left({\bf f}^{\star}\right)^{\sf H}{\bf C}_i{\bf f}^{\star})^2}$
     , then the Hessian matrix $\nabla_{{\bf f}^{\sf H}}^2\gamma({\bf f}^{\star},\lambda)$ is sufficient to be a negative-definite matrix. This completes the proof.
\endproof


\bibliographystyle{IEEEtran}
\bibliography{IEEEabrv,ref_SparseJT_bib}
\end{document}